\newif\ifsubmode
\submodefalse

\newif\ifprintfig
\printfigtrue

\ifsubmode
  \documentclass{aastex}
  \usepackage{epsf}
  \received{} 
  \revised{}
  \accepted{}
  \ccc{}
  \cpright{}{}
\else
  \documentclass[preprint]{aastex}
  \usepackage{epsf}
\fi

\shortauthors{Verdoes Kleijn et al.}
\shorttitle{Core Radio and Optical Emission in nearby FR I radio galaxies}

\newcommand{\etal}{{et al.\ }}

\newcommand{\lta}{\lesssim}
\newcommand{\gta}{\gtrsim}
\renewcommand{\deg}{^{\circ}}

\newcommand{\ergs}{\>{\rm erg}\,{\rm s}^{-1}}
\newcommand{\ergscm}{\>{\rm erg}\,{\rm s}^{-1}\,{\rm cm}^{-2}}
\newcommand{\ergsHz}{\>{\rm erg}\,{\rm s}^{-1}\,{\rm Hz}^{-1}}
\newcommand{\ergscmHz}{\>{\rm erg}\,{\rm s}^{-1}\,{\rm cm}^{-2}\,{\rm Hz}^{-1}}
\newcommand{\ergscmA}{\>{\rm erg}\,{\rm s}^{-1}\,{\rm cm}^{-2}\,{\rm \AA}^{-1}}

\newcommand{\OIII}{[OIII]}

\newcommand{\HalphaNII}{H$\alpha$+[NII]}
\newcommand{\Halpha}{H$\alpha$}

\begin{document}

\title{Core Radio and Optical Emission in the Nuclei of Nearby FR I Radio 
Galaxies\altaffilmark{1}}

\author{Gijs A.\ Verdoes Kleijn,\altaffilmark{2}
Stefi A.\ Baum,\altaffilmark{3} 
P.\ Tim de Zeeuw,\altaffilmark{2} \\
Christopher P.\ O'Dea\altaffilmark{3}
}

\altaffiltext{1}{Based on observations with the NASA/ESA Hubble Space 
       Telescope obtained at the Space Telescope Science Institute, which is 
       operated by the Association of Universities for Research in Astronomy, 
       Incorporated, under NASA contract NAS5-26555.}

\altaffiltext{2}{Sterrewacht Leiden, Postbus 9513, 2300 RA Leiden, 
The Netherlands.}

\altaffiltext{3}{Space Telescope Science Institute, 3700 San Martin Drive, 
Baltimore, MD 21218.}


\ifsubmode\else
\clearpage\fi


\ifsubmode\else
\baselineskip=14pt
\fi


\begin{abstract}
In this paper we analyze the relation between radio, optical continuum
and {\HalphaNII} emission from the cores of a sample of 21 nearby
Fanaroff \& Riley type I galaxies as observed with the VLBA and HST. The
emission arises inside the inner tens of parsec of the galaxies. Core
radio emission is observed in 19/20 galaxies, optical core continuum emission is
detected in 12/21 galaxies and {\HalphaNII} core emission is detected
in 20/21 galaxies. We confirm the recently detected linear correlation
between radio and optical core emission in FR I galaxies and show that
both core emissions also correlate with central {\HalphaNII}
emission. The tight correlations between radio, optical and
{\HalphaNII} core emission constrain the bulk Lorentz factor to
$\gamma \sim 2 -5$ and $\gamma \lta 2$ for a continuous jet and a jet
consisting of discrete blobs, respectively, assuming jet viewing
angles in the range $[30\deg,90\deg]$. Radio and optical core
emissions are likely to be synchrotron radiation from the inner jet,
possibly with a significant contribution from emission by an accretion
disk and/or flow. Elliptical galaxies with LINER nuclei without
large-scale radio jets seem to follow the core emission correlations
found in FR I galaxies. This suggests that the central engines could
be very similar for the two classes of AGNs.
\end{abstract}


\keywords{galaxies: active ---
          galaxies: elliptical and lenticular, cD ---
          galaxies: nuclei ---
          galaxies: structure ---
          ISM: dust, extinction.}

\clearpage


\section{Introduction}
\label{s:intro}

Many galaxies in the nearby universe ($z<0.1$) contain an active
galactic nucleus (AGN) which features a central low-ionization narrow
emission-line region (LINER). These AGNs typically have a bolometric
luminosity $L_{\rm bol} \lta 10^{43} \ergs$ which is orders of
magnitude smaller than those of classical AGN at higher redshifts,
such as Seyferts, Quasars and Broad-line Radio Galaxies. The energy
for the activity is thought to be provided by accretion of matter onto
a supermassive black hole residing at the galaxy center in all AGN
classes. Radio galaxies constitute a subset of the active galaxy
family. These galaxies have radio cores at their centers from which
jets emerge. Radio galaxies are divided in two classes based on
differences in large-scale radio morphology: low-power Fanaroff \&
Riley (1974) type I (FR I) radio galaxies and high-power radio FR II
galaxies (cf.~Bridle \& Perley 1984). Many nearby FR I nuclei show
optical LINER emission (e.g., Baum, Zirbel \& O'Dea 1995). The goal of
this paper is to study the active cores of nearby FR I galaxies and
the relation to their nearby LINER-type counterparts without radio
jets.
 
The hosts of FR I galaxies are almost without exception bright
early-type galaxies (see Ledlow \etal 2001 for counter
example). Quiescent and FR I radio galaxies in nearby Abell clusters
show no statistical differences in their global properties (Ledlow \&
Owen 1995). The two classes do not differ in surface-brightness
profiles, the surface-brightness and size relations, the distribution
over ellipticities, and the occurrence and strength of non-elliptical
isophotes. Zirbel (1993, 1996) found that FR I galaxies are larger
than radio-quiet galaxies at given isophotal magnitude, but this
result was not confirmed by Ledlow
\& Owen (1995). Ledlow \& Owen (1995) did not find differences 
between the radio-loud and quiescent cluster galaxies in the local
density of nearby companions, and in the frequency of morphological
peculiarities or tidal interactions . Studies of samples of FR I
galaxies either in the field or in small groups found indications of
tidal interactions in a majority of them (Gonzalez-Serrano \& Carballo
1993; Colina \& de Juan 1995). It is unknown if this rate is different
for quiescent galaxies in similar environments. In summary, it is
unlikely that host environment and host global properties are decisive
for the capability of an early-type galaxy to become a FR I galaxy,
but tidal interactions might trigger the nuclear activity, at least in
some cases.

With the advent of HST it has become possible to look for host galaxy
differences closer to the AGN itself. In a previous paper (Verdoes
Kleijn \etal 1999) we presented WFPC2 broad- and narrow-band images of
a well-defined complete sample of 21 nearby FR I early-type UGC
galaxies, the `UGC FR I sample'. Similar WFPC2 surveys of radio
galaxies have been carried out also for the 3CR sample (Martel \etal
1999; de Koff \etal 2000) and the B2 sample (Capetti~\etal 2000). The
UGC FR I sample WFPC2 data suggest that the stellar content (i.e.,
color and shape) is very similar to that of quiescent early-type
galaxies including the central few kiloparsec. However, a rigorous
comparison with a quiescent galaxy sample to discern subtle
differences has not yet been performed. Potential AGN fuel is clearly
present in the inner hundreds of parsecs. Central {\HalphaNII}
emission is detected in all sample galaxies and dust is detected in
19/21 galaxies. Other samples of FR I galaxies show emission-line
detection rates of $\gta 80\%$ (e.g., Baum \& Heckman 1989; Morganti
\etal 1992). The detection rate of small-scale dust in quiescent
early-type galaxies is typically only $\sim 40\%$ (van Dokkum \& Franx
1995; Tran~\etal 2001), while emission-line gas is detected in
typically $\sim 60\%$ of the quiescent early-type galaxy population
(Philips \etal 1986; Goudfrooij~\etal 1993). Thus, the detection rate
of potential fuel for an AGN is lower but significant in quiescent
galaxies and hence its presence seems not to be the only factor
determining the on-set of activity.

A VLBA study of the radio properties of the cores and the inner jets
in the UGC FR I nuclei has been presented by Xu \etal (2000). In this
paper we analyze the correlation between this AGN radio emission and
optical continuum and {\HalphaNII} core emission as observed with
HST/WFPC2. We compare the UGC FR I core properties to those of 3CR FR
I cores which have been studied in the radio, optical and X-rays
(e.g., Chiaberge, Capetti \& Celotti 1999; Hardcastle \& Worrall
2000). For the 3CR sample, the correlations between the core emissions
have led to favor the inner jet as their origin. We determine the
validity for different AGN components to produce the core emissions in
the UGC FR I galaxies. We also compare the FR I core properties to
those of nearby LINER-type AGNs without large-scale radio jets.

The outline of this paper is as follows.  Section~\ref{s:data}
describes the sample and data from the literature used in this
analysis. It also describes in detail the optical core flux
measurements.  Section~\ref{s:correlations} quantifies the
correlations between the core fluxes and shows they are not
significantly affected by the ubiquitous central dust
distributions. Section~\ref{s:origincore} describes the arguments
against the existence of compact stellar clusters producing the
optical core emission. We present evidence that the radio and
optical cores are synchrotron emission from the inner jet and
discuss constraints on its Doppler boosting. The contribution to the
core emission by an accretion disk and/or flow is also
considered. Section~\ref{s:halphanii} discusses the possible
excitation mechanisms for the central gas.
Section~\ref{s:fr1sandllagns} surveys the similarities between the FR
I and other LINER cores. Section~\ref{s:summary} summarizes the main
results and final conclusions. Appendix~\ref{a:isophotes} presents the
WFPC2 imaging analysis for UGC 7115 and UGC 12064
(3C449). Appendix~\ref{a:beaming} discusses details of the beaming
models used in Section~\ref{s:boosting}.

Throughout the paper we use a Hubble constant $H_0$= 75
kms$^{-1}$Mpc$^{-1}$.

\section{Sample and Data}
\label{s:data}
The sample of 21 nearby FR I galaxies, which is limited in total radio
flux, was constructed by positional cross-correlation of the Greenbank
1400 MHz sky maps and the UGC catalogue (Condon \& Broderick 1988;
Nilson 1973). The UGC FR I sample and its selection are described in
more detail in Verdoes Kleijn \etal (1999, paper I).

\subsection{Host Galaxy Imaging}
We use the results from the HST/WFPC2 broadband imaging analysis for 19 of
the 21 sample galaxies presented in paper I. In
Appendix~\ref{a:isophotes} we present the same analysis for the two
remaining galaxies, UGC 7115 and UGC 12064 (3C 449). Their data
reduction and isophotal analysis was carried out as described in detail in
paper I.
 
\subsection{Radio Observations} 
\label{s:radioobs}
We use core and total flux densities from a VLA A imaging survey at
1490 MHz (FWHM $1.5''- 3.75''$) (Wrobel, Machalski \& Condon 2001) for
the complete sample and from a VLBA imaging survey at 1670 MHz (FWHM
$\sim 0.01''$) for 18 UGC FR I sample galaxies as presented by Xu
\etal (2000). The VLA A core flux density of 3C 449 is taken from
Feretti \etal (1999). UGC 7115 and 3C 449 lack VLBA observations
because they were added to the sample only after the VLBA observations
were completed. NGC 3801 was not observed with the VLBA because no
core was detected in the earlier VLA observation. Core radio fluxes at
5 GHz (FWHM $\sim 1.4''$ except NGC 4261: FWHM $\sim 15''$) for 11
sample galaxies are taken from Zirbel \& Baum (1995). Total radio
luminosities from the Greenbank 300 ft telescope at 1400 MHz (FWHM
$12'$) for the complete sample are taken from Condon \& Broderick
(1988). Condon \& Broderick corrected for confusion of the 1400 MHz
total flux by nearby sources using VLA C observations (FWHM $18''$)
for 12 of the UGC FR I galaxies. We verified that there was no
confusion for the remaining nine UGC FR I galaxies using the NVSS
catalog (FWHM$\sim 45''$; Condon \etal~1998).

\subsection{Optical Core Emission}
\label{s:opticalcores}
The nuclei of many sample galaxies reveal compact optical core
emission in the WFPC2 $V$- and $I$-band images. To investigate the
nature of these nuclear optical sources (NOS) we determine the
unresolved flux at HST resolution after subtraction of the underlying
galaxy light. For most galaxies we use the WFPC2 images presented in
paper I. For NGC 4486 we use WFPC2 archival short-exposure
observations (F555W and F814W 30s exposures, HST program 5477), since
the observations in paper I have nearly saturated centers. For UGC
7115 and 3C 449 we use the images presented in
Appendix~\ref{a:isophotes}. The WFPC2 point spread function (PSF) has
a FWHM $\sim 0.07''$ at $V$ and $I$ and 85\% of the flux counts, which
we want to isolate from the underlying galaxy light, are contained
inside $r=0.23''$. We approximate the underlying galaxy flux density
inside $r=0.23''$ by a constant. Its value is estimated by measuring
the counts in an annulus with an inner radius of $0.23''$ (i.e., 5 PC
pixels) and a width of $0.0455''$. After subtracting the underlying
galaxy counts from the counts inside $0.23''$, we measure the
resulting flux profile inside this radius. If this profile is
consistent with being unresolved (i.e., FWHM in the range $0.05''-
0.08''$) we decide there is a point source present in addition to the
stellar light. The NOS counts were converted to fluxes using the
`photflam' values given in the HST Handbook (ed. Voit 1998). We
correct the fluxes for Galactic fore-ground extinction and the 15\%
PSF flux which falls outside the $0.23''$ aperture.

Determination of the underlying stellar light can be difficult because
dust is often present and the stellar light profile is sometimes
bright and steep (cf.\ Figure~\ref{f:nosexamples}). In fact, our
procedure tends to overestimate the NOS flux, since galaxies typically
have surface brightness profiles which are not constant but increasing
towards the center. The irregular variations in the ubiquitous central
dust obscuration inhibit accurate estimates of the individual central
surface brightness slopes. Simulations indicate that our procedure
would overestimate the NOS by typically 50\% and at most a factor 2
for the galaxies with a NOS detected in the UGC FR I sample if central
dust were not present. To estimate how the NOS flux estimate depends
on aperture size, we determined the NOS using varying apertures in the
range $0.13''-0.32''$. This typically results in NOS fluxes which
differ by $\sim 30\%$. In addition to stellar galaxy light there is
sometimes, depending on redshift, a contribution to the NOS flux from
the {\OIII}$\lambda\lambda4959,5007$ and {\HalphaNII} emission lines
for the F547W and F555W and F702W filters. Using the central
{\HalphaNII} fluxes from paper I and assuming
{\OIII}/{\HalphaNII}=0.12, typical for the LINER-type spectrum in FR I
galaxies, we estimate that the contribution to the NOS flux in these
filters is typically $\sim 15\%$ and at most $25\%$. Given the
aforementioned uncertainties and the difference in the spatial
resolution of the broad- and narrow-band images, we did not attempt to
correct the NOS for this emission-line contribution. The central dust
might obscure the NOS and stellar background fluxes in several
targets. We lack information on the nature of the dust obscuration to
make a robust correction for internal dust obscuration. As an
indication: if the observed opacity is caused by dust in front of the
stellar flux and the NOS, then the observed $V-I$ WFPC2 colors
typically imply correction factors $\sim 1.5$ for the $V$-band NOS
fluxes. Given all uncertainties, we expect that the detected NOS
fluxes are accurate to within a factor $\sim 2$.

NOS were detected in 12 of the 21 galaxies. We estimate the expected
NOS for the non-detections from the correlation between NOS and radio
core and central {\HalphaNII} emission (see
Section~\ref{s:correlations}). We added the expected NOS counts
artificially to the galaxy image as a scaled PSF to determine the
possible reason for the non-detection. Adding the expected NOS
indicates that it would be easily detected in NGC 741 (which has a
shallow stellar surface brightness profile) and marginally detected in
NGC 541, NGC 4335, NGC 5141, NGC 5490 and NGC 7626. NGC 541, NGC 4335,
NGC 5490 and NGC 7626 have a bright and steep underlying stellar
surface brightness profile which makes detection of a NOS difficult
(cf., Figure~\ref{f:nosexamples}). Thus these non-detections could be
due to observational bias except for NGC 741. In fact, all NOS, except
for UGC 7115, are detected in hosts with relatively shallow inner
stellar surface brightness cusps. In the remaining three galaxies, NGC
3801, NGC 5127 and NGC 7052, the expected NOS could be hidden by
foreground dust for plausible dust-extinction values. In NGC 7052,
there is actually a hint of a very weak blue point source hidden
behind the dust disk. For the galaxies with no obvious obscuration by
foreground dust, the upper limit to the NOS flux was set to the flux
measured within an aperture of $0.13''$ radius (i.e., 80\% of the flux
for an unresolved source) corrected for underlying stellar light. As a
final check of the robustness of the NOS detection procedures, we
verified that both NOS detections and upper limits do not correlate
with the slope of the central stellar background flux profile.

Table~\ref{t:nucfluxdetection} lists the NOS flux detections and upper
limits.  The $V$ and $I$-band NOS fluxes generally differ by less than
$50\%$. Our measurements agree typically within a factor $<2$ with
determinations by Chiaberge, Capetti \& Celotti (1999) for seven overlapping
targets. There is about a factor two disagreement between the F791W
fluxes for UGC 7360 and the F702W fluxes for 3C 449. The former has a
very weak NOS which makes the measurement difficult. Chiaberge \etal
corrected the latter measurement for {\HalphaNII} which we estimated
above to be about 15\% of the NOS flux. This might explain part of the
difference for this measurement.

In conclusion, a NOS is detected in $57\% \pm 16\%$ (21/21) of the FR I nuclei,
dust might hide a NOS in another $14\% \pm 8\%$ (3/21), while upper limits to
the NOS emission can be determined for $29\% \pm 12\%$ (6/21).

\subsection{Central {\HalphaNII} Emission}
\label{s:halphaniidetection}
The WFPC2 narrow-band imaging presented in paper I shows that the
{\HalphaNII} emission is always dominated by a highly peaked nuclear
component. Additional extended emission is detected in several
galaxies. Thus, in close analogy with the unresolved nuclear optical
continuum analysis, we determined the nuclear unresolved {\HalphaNII}
emission using the same procedure. We detected central {\HalphaNII}
flux cores with FWHM $\lta 0.08''$ for the five targets with emission
images on the PC detector and cores with a FWHM $\sim \lta 0.13''$ for
14 of the 15 targets on the WF2 detector. These FWHM indicate that
this emission component is consistent with being unresolved (the
larger PSF FWHM for the WF chip is caused by the larger WF pixel size
($0.0996''$) compared to the PC pixel size ($0.0455''$)). The
{\HalphaNII} flux cores were unresolved at HST resolution in all UGC
FR I galaxies except NGC 3801. For NGC 3801 we set the {\HalphaNII}
flux in the central 9 pixels ($75\%$ of the PSF flux) as an upper
limit. No WFPC2 {\HalphaNII} image is available for UGC 7115. In this
case we estimated the core {\HalphaNII} flux from HST/STIS
measurements, which will be published in a forthcoming paper. The core
{\HalphaNII} fluxes are listed in Table~\ref{t:nucfluxdetection}. From
estimates of the error in the stellar continuum subtraction and in the
core flux measurement procedure we conclude that the core {\HalphaNII}
fluxes are accurate to within 50\%.

\section{Correlations between Core Emissions}
\label{s:correlations}

We now quantify the relation between nuclear radio, optical and
{\HalphaNII} line emission. By definition the optical continuum and
{\HalphaNII} core emission are unresolved at HST resolution (FWHM
$\lta 0.1''$) which corresponds to a region on the order of a few tens
of parsec or smaller. The VLBA cores are unresolved at a resolution of
FWHM $\sim 0.01''$ and originates from a region smaller than a few pc
in size.

\subsection{Radio - {\HalphaNII} Core Correlation} 
\label{s:radiohalpha}
The top two panels of Figure~\ref{f:nucflux} we plot the central
{\HalphaNII} emission as a function of VLBA core peak emission both as
flux and luminosity for 18 sample galaxies. The two quantities are
well correlated: both flux and luminosity correlations have a
significance level $\geq 99.99\%$ using Kendall's tau test (see
Tables~\ref{t:nucflux} and \ref{t:nuclum}). This gives us confidence
that the correlation is not due to selection effects or a common
dependence on distance. 3C274 (M87) is a slight outlier on the
luminosity-luminosity correlation plot. This could be due to the fact
that part of the [NII]6584 emission is redshifted out of the
narrow-band filter (Ford \etal 1994). The alternative of Doppler
boosting of the radio emission is discussed in
Section~\ref{s:accretion}. Two of the three missing objects, UGC 7115
and 3C 449, lack VLBA observations for reasons unrelated to their
properties (see Section~\ref{s:radioobs}). No radio core was detected
in NGC 3801. NGC 3801 has no detected core {\HalphaNII} emission. Thus
we do not expect the missing objects to alter the correlations
significantly. The luminosity-luminosity correlation extends over
approximately two orders of magnitude in radio and {\HalphaNII}
luminosity. A linear regression fit to the logarithmic values of the
luminosities shows that the {\HalphaNII} luminosity is roughly
proportional to the the radio core luminosity to the power 0.6
(Table~\ref{t:nuclum}).

How does this correlation compare with previous studies of FR I
nuclei?  Baum, Zirbel \& O'Dea (1995; see also Zirbel \& Baum 1995,
ZB95) performed an extensive study of a large sample of $\sim 70$ FR I
galaxies, with radio and emission-line measurements gathered from the
literature. Their FR I sample extends $\sim$ 1 magnitude brighter in
host magnitude and total and core radio luminosities. ZB95 found only
a weak luminosity-luminosity correlation between integrated
{\HalphaNII} emission and the radio core at 5 GHz (FWHM $\sim 1.4''$)
for FR I galaxies. They found a much stronger correlation between
integrated {\HalphaNII} emission and host magnitude. They also found a
correlation with total radio luminosity, which turns out to be a weak
(2$\sigma$) correlation after taking into account the correlation with
host magnitude. ZB95 concluded that the {\HalphaNII} emission gas in
FR I galaxies was most likely predominantly excited by radiation from
old stars. Table~\ref{t:nuclum} lists the parameters of the linear
regression fits to the correlation between {\HalphaNII} emission and
host optical magnitude and total radio power for the UGC FR I
sample. The fits agree with those obtained by ZB95 after correction
for the difference in the assumed value of the Hubble constant and
difference in radio frequency. However, for the UGC FR I sample the
resulting correlations are much weaker than the correlation between
{\HalphaNII} and radio core. We compare the {\HalphaNII} luminosities
for seven overlapping sources with {\HalphaNII} flux measurements in
the UGC FR I and ZB95 sample. For four sources ZB95 report larger
{\HalphaNII} luminosities. For the other three sources ZB95 actually
use smaller {\HalphaNII} luminosities. However, from examination of
the observational details there is clear indication for considerable
uncertainty in each of these three measurements. Thus, the difference
in correlation strength seems caused by the fact that we consider the
core {\HalphaNII} fluxes, while ZB95 considered total fluxes, which
might be dominated by an extended component. The correlation between
the central {\HalphaNII} emission and radio core emission is also
present if we restrict the sample to the 11 overlapping targets and
use the 5GHz core emission values from ZB95 (see
Tables~\ref{t:nucflux} and
\ref{t:nuclum}). Similarly, there is a good correlation between the
central {\HalphaNII} emission and the VLA core emission (FWHM $\sim
1.5'' - 3.75''$). It is significant at $>99.9\%$ level and has a
similar slope as the correlation with VLBA core emission
(Tables~\ref{t:nucflux} and~\ref{t:nuclum}). Thus the difference in
spatial resolution of the VLBA, VLA and 5 GHz measurements does not
seem relevant. The correlation with extended VLBA radio emission is
significant at the $>99\%$ level and has a similar slope. No strong
correlation is found between the extended VLA emission and the
{\HalphaNII} emission.

Thus not surprisingly, the AGN, which produces the radio core emission
(see also discussion in Section~\ref{s:jet}), very likely excites the
gas in the central few hundreds of parsec in the FR I nuclei. More
interestingly, the excitation energy correlates tightly with the
central radio power. The correlation between total {\HalphaNII}
luminosity and host galaxy magnitude, as found by ZB95, probably
indicates that the gas on kpc scales is excited by some other
process(es) (e.g., radiation from old stars).

\subsection{Radio - Optical Core Correlation}
\label{s:radionos}
The middle two panels in Figure~\ref{f:nucflux} plot the nuclear
optical source emission in the $I$-band as a function of VLBA core
peak emission both in flux and luminosity for 16 of the 21 sample
galaxies. In six galaxies the NOS measurement is an upper limit. The
two quantities are well correlated: both flux and luminosity
correlations have a significance level larger than 99.9\% using
Kendall's tau test, which takes the upper limits into account (see
Tables~\ref{t:nucflux} and \ref {t:nuclum}). This gives us confidence
that the correlation is not due to selection effects or a common
dependence on distance. Figure~\ref{f:nucflux} shows the correlation
with the $I$-band NOS, because the $V$-band NOS can be more affected
by dust obscuration. Both NOS emissions correlate well with the VLBA
radio core (see Tables~\ref{t:nucflux} and~\ref{t:nuclum}). Two of the
five missing galaxies, UGC 7115 and 3C 449 both host a NOS (see
Table~\ref{t:nucflux}) but lack VLBA observations for reasons
unrelated to their properties (see Section~\ref{s:radioobs}). No NOS
is detected in NGC 3801, NGC 5127 and NGC 7052, which could be due to
large foreground dust obscuration
(cf.~Section~\ref{s:opticalcores}). Thus all missing objects might
host a NOS in principle. There is no a priori reason to believe that
the missing targets will alter the correlation significantly. For five
of the six upper limits, the NOS flux, as expected from the radio -
optical core correlation, would be difficult to detect (cf.\
Section~\ref{s:opticalcores}). The slope of the correlation between
the logarithm of the VLBA core and the $I$-band optical core
luminosities is $1.18 \pm 0.25$ (see Table~\ref{t:nuclum}). The slope
for the $V$-band optical core is less constrained but slightly
steeper: $1.47 \pm 0.35$.  Thus the majority of the 21 FR I nuclei do
show NOS emission, which is roughly proportional to the radio core
emission.

Chiaberge, Capetti \& Celotti (1999) used HST/WFPC2 observations to
determine optical core emission in all 33 FR I galaxies in the 3CR
catalogue. They detected a NOS in $69\% \pm 15\%$ of the galaxies,
dust could hide a NOS in $13\% \pm 6\%$, and no NOS was detected in
$19\% \pm 8\%$. Thus, the detection rates for the 3CR FR I sample and
UGC FR I sample are similar. Their NOS measurements agree well with
our results for the seven overlapping 3CR galaxies (see
Section~\ref{s:opticalcores}). Chiaberge, Capetti \& Celotti also
found a linear correlation between radio and optical core
flux. Figure~\ref{f:nocradiocore} show the $I$-band NOS versus radio
core flux for the 16 UGC galaxies and the 20 3CR galaxies which are
not part of the UGC FR I sample. The radio core flux at 5 GHz is
plotted for the galaxies taken from Chiaberge, Capetti \& Celotti, as
listed in their paper. For the UGC FR I sample, the VLA radio core
flux at 1.49 GHz is actually plotted. The 5 GHz and 1.49 GHz
observations have similar beamwidths. The measurements can therefore
be compared directly under the assumption of flat spectral
slopes. This seems to be a reasonable assumption because the fluxes in
the UGC FR I galaxies with both 5GHz and 1.49 GHz observations
typically differ by a factor $<2$.  The combined sample of 37 FR I
galaxies suggests a single relation between radio and optical core
emission. The only outlier is 3C 386. As discussed by Chiaberge,
Capetti \& Celotti (1999), 3C 386 displays broad {\Halpha} emission
lines which is highly unusual for a FR I galaxy. We performed a
linear regression fit using Kaplan-Meier residuals on the combined
sample (Isobe, Feigelson \& Nelson 1986). The outlier 3C386 is
excluded from the fit. Both 3C 28 and 3C 314.1 are excluded from the
fit as well because they have upper limits to both their radio and
optical flux, which cannot be dealt with by the fitting routine. The
linear fit to the logarithms of optical core flux as a function of
radio core flux is significant at the $>99.99\%$ level and has a slope
of $1.0 \pm 0.1$.

Hence, both the 3CR and UGC sample of FR I galaxies indicate a linear
relation between radio and optical core emission. The proportionality
makes a common origin for the radio and optical core flux plausible.

\subsection{Optical - {\HalphaNII} Core Correlation}
\label{s:noshalpha}
We should expect a correlation between nuclear optical core and
central {\HalphaNII} emission, given the correlations between the radio
core and {\HalphaNII} emission and between the radio and optical
core. Any correlation, if present, might show considerable scatter due
to the larger errors in optical core and {\HalphaNII} emission in
comparison with the error in radio core flux.

The central {\HalphaNII} emission as a function of NOS emission is
shown in the bottom panels of Figure~\ref{f:nucflux} for 18 of the 21
galaxies. The three galaxies missing from the plot; the cores of NGC
3801, NGC 5127 and NGC 7052 are too obscured by dust to determine the
NOS flux (cf. Section~\ref{s:opticalcores}). It is not expected that
the missing galaxies bias the relation between {\HalphaNII} and NOS
emission. The optical core and central {\HalphaNII} emission are
indeed correlated. The fluxes and luminosities are correlated at a
significance larger than $99\%$, using Kendall's tau test (see
Tables~\ref{t:nucflux} and
\ref{t:nuclum}). There are two outliers from the general relation:
NGC 4261 and NGC 2329. In NGC 4261 the NOS is detected at the bottom
of a dip in the central luminosity profile, which is due to dust
obscuration. Thus, the intrinsic NOS flux might be severely
dimmed. For NGC 2329 there is no clear indication that its NOS and/or
{\HalphaNII} emission flux is biased. However, in paper I we noted a
slight linear extension of the nuclear optical core. Possibly, an
extra optical component (e.g., slightly extended optical jet) in NGC
2329 causes an overestimate of the NOS flux. The {\HalphaNII} and
optical emission measurements have similar errors. Neither of the
quantities can be considered an independent variable in a linear
regression fit. We perform a linear least squares fit which takes into
account the error in both optical core flux and {\HalphaNII} flux
(Press \etal 1992). The method cannot take into account the data with
upper limits, which however suggest a similar relation as the
detections. The best fit to the logarithms of the {\HalphaNII}
luminosity as a function of the NOS luminosity has a slope $0.55 \pm
0.17$ which is similar to the slope between {\HalphaNII} and radio
core emission (Table~\ref{t:nuclum}).

\subsection{Correlation between Regression Fit Residuals}
\label{s:residuals}
There are three pairs of linear regression fits, each one with one of
the three core luminosities as the independent variable and fitting
the other two. We determined if a correlation exists between the
residuals of each pair of fits. We see a trend instead of a scatter
plot only if the central {\HalphaNII} luminosity is taken as the
independent variable do (Figure~\ref{f:residuals}). This is additional
suppport for the view that the optical and radio core emission share a
common origin.

\subsection{Dust Obscuration}
\label{s:obscuration}
Paper I discusses the dichotomy in apparent dust morphology: (i) dust
disks, which appear as smooth ellipses and (ii) dust lanes, which are
irregular dust filaments. This extended central dust, detected in 19
of the 21 UGC FR I galaxies, might potentially obscure a significant
fraction of the central optical line and continuum emission.

As a general check if dust affects the measurements of the optical
continuum and {\HalphaNII} core emissions we checked for any
correlation between the two core luminosities and IRAS far infrared
luminosities which trace the extended hot dust (possibly with
additional emission from stars) at 12 and 25 micron and cool dust at
60 and 100 micron (e.g., Knapp, Bies \& van Gorkom 1990). IRAS flux
measurements are available for 17 UGC FR I galaxies (Knapp \etal
1989). About half of those are flux upperlimits. No correlation or
trend is observed for the {\HalphaNII} and optical core luminosities
and any of the IRAS luminosities (see Table~\ref{t:nuclum}).

Another possibility is that the obscuration depends on the orientation
of the central dust. For example, the amount of obscuration along the
line of sight increases going from face-on disks to edge-on disks
and/or dust lanes. Could this induce the correlation between
{\HalphaNII} and optical core luminosity? Figure~\ref{f:nucflux} shows
that a large but not too unreasonable range in $V$-band dust opacity
of $A_V \sim 5$ tends to spread the observed luminosities roughly
along the observed correlation, if the intrinsic luminosities were
quite similar among the sample. However, there is no correlation
between either {\HalphaNII} or optical core luminosity and dust-disk
inclination (Table~\ref{t:nuclum}; an inclination for the disks was
derived by assuming intrinsically round disks, cf.\ paper I). The
average {\HalphaNII} luminosity is slightly lower ($\sim 0.5$ dex) in
dust-lane galaxies than in dust-disk galaxies. Dust lanes might
intersect the line of sight towards the {\HalphaNII} emission, while
relatively face-on dust disks might not, or there might be an
intrinsic difference in {\HalphaNII} luminosities of dust-disk and
dust-lane galaxies. There is also no indication of significant
foreground dust obscuration from the $V-I$ color of the optical core
emission as a function of dust-disk inclination. Only a statistically
insignificant trend is present (Table~\ref{t:nuclum}), for which the
slope is too shallow to explain the observed range in
luminosities. Thus, foreground dust does not appear to systematically
affect the {\HalphaNII} and optical core emission. This makes it very
unlikely as well, that dust obscuration induces the correlations of
{\HalphaNII} and optical core emission as a function of radio core
emission, which extend over two decades or more in luminosity.

Could foreground dust obscuration cause the scatter in the
correlations of optical core and {\HalphaNII} luminosity as a function
of radio core luminosity? The observed scatter of about one decade in
both correlations requires a reasonable optical depth variation of
$A_V \sim 3$ between the least and most obscured cases in the UGC FR I
sample (see arrow in Figure~\ref{f:nucflux}). However, there is no
trend between the residuals of the linear regression fits as a
function of dust-disk inclination (Table~\ref{t:nuclum}). There is
also no systematic difference between dust-disk and dust-lane
galaxies.

\subsection{Summary Core Emission Correlations}
The conclusion from Section~\ref{s:correlations} is that the radio,
optical and {\HalphaNII} core emission are all mutually well
correlated. The analysis indicates that the optical core emission is
proportional to the radio core emission. The central {\HalphaNII}
emission is consistent with being roughly proportional to the square
root of both optical and radio core emission. It is highly unlikely
that variations in foreground dust obscuration cause the correlations
between {\HalphaNII}, optical and radio core emission. These
correlations rather reflect physical connections between the three
emissions. Foreground dust is also unlikely to induce the scatter in
the correlations. We will show in Section~\ref{s:boosting} that
Doppler boosting is a more viable explanation for the scatter (which
probably contains an additional contribution from intrinsic scatter,
flux variability and measurement errors as well).

\section{Origin of the Optical and Radio Core Emission}
\label{s:origincore}

We can use the observed correlations between the core emissions to
constrain the origin of these emissions. We will consider the
possibility that the optical core emission is produced by a compact
nuclear stellar cluster. This will turn out to be implausible and we
then go on to consider the possibility that the optical and radio core
emissions originate in the unresolved inner jet and/or accretion disk
or flow.
 
\subsection{Compact Young Stellar Clusters?}
\label{s:cluster}
There are two main methods to discriminate between a stellar and
non-stellar origin for the optical core emission. First, sufficiently
strong nuclear stellar absorption lines will indicate a stellar origin
for the optical core emission. Second, the nuclear gas emission-line
ratios can be used to distinguish between excitation by an AGN or
stars. A spectral analysis of the optical core emission has been
performed for the two nearest members of the UGC FR I sample, M84 and
M87, at sufficient spatial resolution to isolate it from the stellar
galaxy background . Bower~\etal (2000) conclude that the nuclear
continuum source in M84 is an AGN from the analysis of an HST/STIS
optical spectrum (covering $\sim 2900$ {\AA} - 6800 {\AA}), taking
into account the nuclear dust obscuration. The observed emission-line
ratios exclude excitation by young massive OB stars, while the
weakness of the stellar absorption features exclude a cluster of A
stars. Last but not least, the variability of the optical core (75\%
in the $V$-band over a five year period) argues against a stellar
origin for the optical core in M84. Dressler \& Richstone (1990) and
Kormendy (1992) argued that the nucleus in M87 has a non-thermal
origin from the weakness of stellar absorption lines in the $B$- and
$V$-band. Carter \& Jenkins (1992) and van der Marel (1994) reached
the same conclusion using ground-based optical and near-IR
spectra. Both studies did not confirm the earlier detection of strong
stellar Ca II triplet lines by Jarvis \& Melnick (1991). Tsvetanov
\etal (1999) determined from wide-band ($\sim 1600$ {\AA} - 8000
{\AA}) HST/FOS peak-up optical images that also the core of M87 varies
by a factor $\sim 2$ over 2.5 months, a factor $\sim 0.25$ over three
weeks and a factor $<0.025$ over one day, which again strongly argues
against a stellar origin for the optical core emission. Both M84 and
M87 have optical core luminosities representative of the UGC FR I
sample.  Thus for two typical members of the 21 UGC FR I galaxies it
has been convincingly shown that the optical core emission source is
not a stellar cluster but must be AGN related.

It is possible that (some of) the 19 remaining UGC FR I nuclei do host
a nuclear stellar cluster. In fact, such compact nuclear stellar
clusters, with sizes smaller than a few tens of parsec, have been
found in other galaxies. Maoz \etal (1998) performed a UV
spectroscopic study of the compact nuclei of a sample of seven nearby
UV-bright galaxies with LINER type emission using HST (cf.\
Table~\ref{t:liners}). Their nuclear SED appear similar to the UGC FR
I galaxies, which also show LINER emission and compact UV
emission (Ho, 1999b; Ford
\etal 1994; Ferrarese~\etal 1996; van der Marel \& van den Bosch 1998; Zirbel~\& Baum 1998). 
All seven UV nuclei from the LINER sample are compact and four are
consistent with being unresolved, which implies scales of a few parsec
(Maoz \etal 1995; Barth \etal 1998). Maoz \etal (1998) detected UV
stellar absorption-lines in NGC 404, NGC 4569, NGC 5055 and possibly
NGC 6500 and concluded that the dominant UV continuum source is a
cluster of young massive stars in these galaxies. An AGN is most
likely the UV continuum source in NGC 3031 and NGC 4579, which show broad
UV emission lines and in NGC 4594, which has narrow UV emission lines
(cf.\ Nicholson \etal 1998). We will refer to the former four galaxies
as stellar-type LINERS and to the latter three as AGN-type LINERS.

We reduced HST/WFPC2 archival optical images for the LINER sample
to study the optical cores (see Table~\ref{t:liners}). We determined
the unresolved NOS flux in six of the seven galaxies in the
same manner as done for the UGC FR I galaxies
(cf.~Section~\ref{s:opticalcores}).  Unfortunately, the core of NGC
5055 was saturated. We find an unresolved optical core in all six
LINER galaxies except NGC 6500. The same result was derived for the UV
cores, using detailed PSF modeling. In NGC 4569 the optical core was
unresolved in the $V$-band but slightly resolved in the $I$-band. As
opposed to Maoz \etal (1995), Barth~\etal (1998) also detect
slightly resolved UV emission in NGC 4569, which however is strongly
peaked towards the nucleus. Our NOS fluxes agree within a factor of
$\sim 2$ with measurements by Ho (1999b) for three overlapping
targets. The agreements in the spatial extent and flux measurement of
the nuclear optical emission support the idea that our simple method
for detecting and measuring unresolved emission is accurate for our
purposes. Table~\ref{t:liners} lists the optical core fluxes and
central {\HalphaNII} and 1.4 GHz radio luminosities for the LINER
sample gathered from the literature.

Figure~\ref{f:liners} shows again the correlations between the central
luminosities for the UGC FR I sample but now with the LINER sample
data overplotted. The AGN-type LINERS and UGC FR I galaxies have
similar NOS and {\HalphaNII} luminosities. The LINER {\HalphaNII}
luminosities are measured in apertures of typically a few arcsecond
radius and hence overestimate the core {\HalphaNII} luminosity. Taking
this into consideration, the AGN-type LINERS seem roughly to extend
the core emission relations of the UGC FR I sample. The stellar-type
LINERS also have similar NOS luminosities. NGC 4569 has a comparable
{\HalphaNII} luminosity as well, but NGC 404 has an order of magnitude
lower {\HalphaNII} luminosity. Furthermore, the upper limits to the
radio core luminosities of the stellar-type LINERS are significantly
below those of UGC FR I galaxies except for one (NGC 6500). Thus, in
general the AGN-type LINER nuclei resemble the UGC FR I nuclei in
their core emission properties more closely than the stellar-type
LINERS. It is especially the radio core luminosity that separates the
stellar-type LINERS from the AGN-type LINERS and UGC FR I
galaxies. The core radio emission in the UGC FR I nuclei is certainly
non-thermal AGN emission (cf.~Section~\ref{s:jet}). The resemblance
with the AGN-type LINERS suggest an AGN source for the optical
continuum and line emission in all UGC FR I nuclei is more likely than
a stellar origin. It is unlikely that systematic differences in
obscuration by central dust, which is is commonly present in both
stellar- and AGN-type LINER and UGC FR I galaxies, conspire to produce
the overlap and separation of the nuclear luminosities. Indeed, the
$V-I$ colors of the LINER cores (only available for NGC 4579 and NGC
4594) and the UGC FR I cores are similar.

We conclude that a stellar origin for the optical core emission is highly
unlikely for two reasons. First, previous spectral studies have
directly excluded a nuclear stellar cluster as the source of the
optical core emission for two typical members of the UGC FR I
sample. Second, the UGC FR I nuclei resemble the nuclei of AGN-type
LINERS more closely than stellar-type LINERS in their radio and
optical core and central {\HalphaNII} luminosities. In contrast, an
AGN origin for the optical cores in all UGC FR I galaxies is strongly
favored by the tight correlation with radio core emission, which is
certainly produced by the AGN.

\subsection{Jet Synchrotron Emission}
\label{s:jet}
We shift our attention to a scenario in which the radio and optical
core emission originate in the inner jet. The radio core emission in
radio galaxies is generally thought to be self-absorbed non-thermal
synchrotron emission because it has a high degree of polarization, a
high brightness temperature and typically a flat spectral slope at
radio wavelengths (cf.~Krolik 1999 for a general discussion). The
tight linear relation between radio and optical core flux (see
Section~\ref{s:radionos}) then suggests that the optical core
emission is also jet synchrotron emission. It is therefore useful to
compare the cores to extended optical jets, in which both radio and
optical emission are commonly thought to be synchrotron emission
(e.g., Lara \etal 1999). The radio-to-optical spectral index
$\alpha_{\rm ro}$ (defined by $S_{\nu} \sim
\nu^{-\alpha}$) of the UGC FR I cores varies between 0.44 and 0.86 using any combination of
either $V$- or $I$-band NOS and either VLA or VLBA core fluxes. These
values are consistent with those found for extended optical jets in
the UGC FR I sample: 3C 66B, 3C 264 and M87 (Butcher, van Breugel \&
Miley, 1980; Biretta \etal 1991; Lara \etal 1999). The spectral index
might be affected by core flux variability, because the optical and
radio observations were made at different epochs.  The optical cores
in M84 and M87 are known to vary by a factor of a few (see
Section~\ref{s:cluster}). Not much is known about the flux variability
of radio cores in FR I galaxies, but a factor of a few seems
reasonable (Shukla \& Stoner 1996). As an indication, a flux ratio
variation of a factor of 10 results in $\alpha_{\rm ro}$ variation of
$\sim 0.2$. Although the $V$- and $I$-band observations were made at
the same epoch, the $V$-to- $I$ spectral index, $\alpha_{VI}$, is much
less secure than $\alpha_{\rm ro}$, due to the large relative errors
of the optical core fluxes and the shorter frequency range (the error
in $\alpha_{VI}$ is approximately 6 times the error in the logarithm
of the flux ratio). We determined $\alpha_{VI}$ taking into account
the filter passbands (using {\tt calcphot} in the IRAF {\tt synphot}
package). Extended optical jets have been detected in three UGC FR I
galaxies: 3C 66B, 3C 264 and 3C 274. The optical spectral index of the
core emission is very similar to that of the extended jet emission in
3C 264 and M87. The optical jet in 3C 66B is too faint to make a
reliable comparison. For the complete sample of UGC FR I cores, we
determine $\alpha_{VI} \sim 1-2$ typically, which is again in the same
range as found for extended optical jets (e.g., Crane \etal 1993 and
references therein; Biretta
\etal 1991; Martel \etal 1998). However, steeper spectral indices up
to $\alpha_{VI} \sim 5$ are observed as well. In addition to the large
measurement uncertainty, any foreground dust (either present in our
Galaxy or the galaxy itself) will steepen the observed optical core
spectral index. For example, if one uses the spectral flux density at
5500 {\AA} and 8000 {\AA} to compute the spectral index and assume
foreground dust with Galactic reddening law then $\alpha_{\rm obs} =
\alpha_{\rm int} + 0.2A_V / \log (\nu_V /
\nu_I) \approx \alpha_{\rm int}+1.2A_V$. Hence 
extinction from foreground dust could account for the steeper optical
spectral indices.

Thus two conclusions can be drawn. First, the tight linear correlation
between optical and radio core flux strongly suggests that also the
optical cores are synchrotron emission. Second, the similarities in
radio and optical emission from extended jets are consistent with both core
emissions originating in an unresolved inner jet.

\subsection{Doppler Boosting of Core Emission}
\label{s:boosting}
If indeed radio and optical core emission originate in the inner jet
both continuum emissions could be significantly Doppler boosted. This is
in contrast with the {\HalphaNII} core emission, which is very likely
not Doppler boosted. We will assume in the following that the
{\HalphaNII} flux is emitted isotropically and not affected by dust
obscuration, based on the results in Section~\ref{s:obscuration}. We
will further assume that the radio and optical core emission originates
from a symmetric jet. The logarithmic factor ${\cal B}$ by which the
core flux is (de-)boosted is then (cf.~Appendix~\ref{a:beaming})
\begin{equation}
\begin{array}{c}
{\cal B} = \log \frac{\delta^p(\theta)+\delta^p(\pi+\theta)}{2}, \\
\delta(\theta) = [\gamma(1-\beta\cos\theta)]^{-1}, \\
\end{array}
\end{equation}
where $\gamma$ is the bulk Lorentz factor, $\beta$ the jet velocity in
units of the velocity of light, $\theta$ the jet viewing angle and $p$
the jet structural parameter. The value of $p$ for FR I jets is not
known. We will consider two cases: (i) a continuous jet isotropically
emitting in the source rest-frame (i.e., $p=2+\alpha$) and (ii) a jet
consisting of discrete blobs isotropically emitting in the source
rest-frame (i.e., $p=3+\alpha$) (see e.g., Urry \& Padovani 1995). For
simplicity we will assume a flat core radio spectrum in the source
rest frame (i.e., $\alpha=0$).

In Appendix~\ref{a:beaming} we determine the effect of Doppler
boosting on the observed correlation of fluxes which are intrinsically
related by a power law. The main effect is to introduce scatter in the
observed relationship. The scatter increases with $\gamma$ for a given
$p$ and given range in jet viewing angles. We will first infer
$\gamma$ assuming that the viewing angles, $\theta$, are randomly
spherically distributed in the range $[30\deg,90\deg]$, which is
plausible for the UGC FR I jets (see also
Section~\ref{s:accretion}). We then discuss the critical dependence of
the results on the assumed range in viewing angles.

Thus we can infer a typical $\gamma$ for the sample from the observed
standard deviation of $\sim 0.3$ for the linear regression fit of the
VLBA radio core flux as a function of {\HalphaNII} flux in log-log
space. Figure~\ref{f:boosting}a plots predicted standard deviations as
a function of a constant bulk Lorentz factor $\gamma$ for $p=2$ and
$p=3$. For the continuous jet the observed standard deviation
corresponds to a typical bulk Lorentz factor $\gamma \approx 2.7$ (for
$\theta=[30\deg,90\deg]$). For a jet consisting of discrete blobs the
observed standard deviation corresponds to $\gamma \approx 1.5$. The
linear regression fits to the VLA core or VLBA extended emission
instead of the VLBA core emission also have standard deviations $\sim
0.3$. Hence similar constraints on $p$ and $\gamma$ are derived for
the radio emission at these scales. The scatter in the optical -
{\HalphaNII} core correlation has a standard deviation $\sim
0.5$. This larger value (compared to the radio core correlation) could
be consistent with the same $\gamma$ as inferred for the radio cores
if it is caused by the larger relative measurement error of the
optical core emission. Indeed, the observed standard deviations are
probably only partially due to Doppler boosting, because intrinsic
scatter, flux measurement errors and perhaps other effects contribute
to the observed scatter. In fact the mean error in the central
{\HalphaNII} flux is $\log 1.5 \approx 0.2$. Thus any inferred
$\gamma$ is likely to be an upper limit (cf.\
Figure~\ref{f:boosting}a).

Another constraint on $\gamma$ can be derived from the observed range
in residuals. The observed range in residuals is $\sim [-0.5,1.0]$
(cf.~Figure~\ref{f:residuals}). Figure~\ref{f:boosting}b shows the
predicted range in residuals for different $p$ and $\gamma$. For the
continuous jet ($p=2$) the observed range is reproduced for $\gamma
\approx 5$, while for the jet consisting of discrete blobs ($p=3$) the
observed range is reproduced for $\gamma \approx 2$. Interestingly the
predicted asymmetry around 0 in the spread of the residuals is present
in the observations as well.

Thus, for $p=2$ the observed range in residuals suggests a significant
spread around the typical Lorentz bulk factor $\gamma = 2.7$ derived
from the spread of the sample as whole with some jets reaching $\gamma
= 5$ at least. For $p=3$ both the total range and dispersion of the
residuals is consistent with $\gamma \lta 2$.

There is further circumstantial evidence that the spread around the
linear regressions is indeed caused by Doppler boosting. On the one
hand, Xu~\etal (2000) note that NGC 2892, NGC 4335, NGC 5127 and
possibly NGC 7626 do not have extended structures associated with the
VLBA radio cores and that their large scale jets are more symmetric
than those in the other UGC FR I galaxies. As argued by Xu et al.,
a possible scenario is that these jets lie close to the plane of the
sky and hence have VLBA jets deboosted below the detection
threshold. At larger scales the jets are slowed down and hence have
negligible Doppler boosting. On the other hand, there is evidence that
extended optical jets are only detected when the jet angle is
sufficiently close to the line of sight (Sparks~\etal 1995;
Sparks~\etal 2000). Extended optical jet emission has been detected in
three UGC FR I sample galaxies: 3C66B, 3C264 and
M87. Figure~\ref{f:boostingc} plots the residuals as a function of radio
and optical core luminosity. For the radio residuals, all the galaxies
without extended VLBA structures have negative residuals while all
galaxies with extended optical jets have positive or close to positive
residuals. There is also a trend for more luminous cores to have more
positive residuals. This is expected if the increase in core
luminosity is (at least partially) caused by Doppler boosting. The
corresponding plot for the optical core emission is consistent with
these results. It shows more scatter, which is expected as the errors
in the optical core flux are larger than those of the radio core
flux. Finally, the correlation between the residuals of the
optical/radio core correlations with {\HalphaNII} emission suggest
they are Doppler boosted as well (cf.~Figure~\ref{f:residuals}). The trends
are mainly driven by the few points at the high and low end of the
distributions and hence larger samples are required to confirm these
results.

The bulk Lorentz factors inferred above depend on the assumptions of
$\alpha=0$ and the range in jet viewing angles
[$\theta_0,\theta_1$]. Our model depends on $\alpha$ only through
$p=p_0+\alpha$, where $p_0={2,3}$.  Thus varying $\alpha$ is identical
to varying $p$, which is discussed above. The results depend
critically on the assumed minimum jet angle $\theta_0$ because Doppler
boosting increases rapidly as a function of $\theta$ at small $\theta$
(Figure~\ref{f:boosting}b). Figure~\ref{f:boosting} illustrates the
effect of varying $\theta_0$ between $[20\deg,40\deg]$.

BL Lac galaxies are commonly thought to be a subset of FR I galaxies
where the jet is pointed very close to the line of sight (e.g., Browne
1983; Urry
\& Padovani 1995). This FR I-BL Lac unification scheme typically
requires larger bulk Lorentz factors, $\gamma > 5$, than inferred by
our analysis. In general, BL Lac jet studies infer a larger $\gamma$
than FR I jet studies (Urry \& Padovani 1995). Two-layer jets have
been invoked to reconcile this aparent discrepancy. A fast inner spine
is surrounded by a slower outer layer with. In FR I galaxies the jet
emission is dominated by the slow outer layer and in BL Lac galaxies
by the fast spine. For example, Chiaberge \etal (2000) argue for a
two-velocity jet, based on the core optical, radio and X-ray fluxes
assuming that these fluxes are dominated by jet emission. They infer a
fast inner spine with $\gamma \sim 15- 20$ surrounded by a slower
outer layer with $ 1 < \gamma \lta 2$. These results are derived using
jet de-beaming models of BL Lac SEDs for $p=3$. So this two-velocity
component jet model infers the same $\gamma \lta 2$ as derived by us
from the correlation between radio core flux and central {\HalphaNII}
flux for our UGC FR I sample for $p=3$ and
$\theta=[30\deg,90\deg]$. Additional evidence for the presence of
two-velocity jets comes from detailed modeling of extended FR I jets (
e.g., Owen, Hardee \& Cornwell 1989; Komissarov 1990; Hardcastle \etal 1996; Laing \etal
1999). Giovannini \etal (2001) use the scatter in the correlation
between VLBI core and total radio power for a sample of 13 FR I, 6 FR
II and 8 compact radio objects to derive $\gamma \sim 3 - 10$ for the
cores under the assumption of $p=2$. The $\gamma \sim 2 - 5$ derived
by our method is roughly consistent with this, especially since lower
bulk Lorentz factors are typically expected in FR Is compared to FR
IIs.

Thus our constraints on $\gamma$ from the tight correlation between
radio core and {\HalphaNII} emission agree with constraints on
$\gamma$ from other methods. To establish whether cores are mildly or
non-relativistic (i.e., $\gamma \lta 2$) or highly relativistic (i.e.,
$\gamma > 2$) requires precise knowledge of the jet structure since
the inferred value of $\gamma$ of the core emission critically depends
on the value of $p$ within the typically assumed range $p=2-3$ and the
range in jet viewing angles and how much of the scatter is due to
non-Doppler boosting effects.

\subsection{Accretion flow emission}
\label{s:accretion}
The absence of compelling evidence for bulk Doppler boosting with
$\gamma > 2$ leaves open the possibility that a component, different
from the inner jet, dominates the radio and optical core emission. For
example, it could be synchrotron (and/or inverse Compton) emission
from the accretion disk and flow or wind. The inferred bolometric AGN
luminosities of LINER and UGC FR I sample galaxies are three to five
orders of magnitude below the Eddington luminosities (Ho 1999b). They
are therefore more easily explained by advection dominated accretion
flows (ADAF) than standard, geometrically thin, optically thick
accretion disks. In these models the radio emission is synchrotron
emission and the optical emission is synchrotron and/or inverse
Compton emission (e.g., Narayan, Mahadevan \& Quataert 1998).
Quataert, Di Matteo \& Narayan (1999) obtain a reasonable fit to the
central radio to X-ray spectral energy distribution (SED) of two
AGN-type LINER galaxies NGC 3031 and NGC 4579, which have unresolved
nuclear cores with SEDs similar to the UGC FR I galaxies (cf.\
Section~\ref{s:cluster}). The authors use a combination of a standard
thin accretion disk with an ADAF model for the inner accretion
disk. However, the best-fit model critically depends on the amount of
dust obscuration for the optical and UV fluxes. For the core of M87,
an ADAF model by Reynolds \etal (1996) underpredicts the optical flux
by a factor of $>100$. More recently, Di Matteo \etal (2000) obtain a
reasonable fit to the radio to X-ray SED of M87 using an ADAF model
combined with an outflow. The models that fit the observed X-ray SED
underpredict the nuclear optical emission by a factor of $\sim 5$. Di
Matteo \etal note evidence for nuclear synchrotron jet emission from
the radio and millimeter fluxes. Perhaps the X-ray emission also
contains a beamed component. Yi \& Boughn (1999) show that ADAF models
predict within a factor of a few the radio core fluxes in NGC 4261,
NGC 4374 and M87.

Thus, ADAF type models seem to provide a reasonable global fit to the
nuclear SEDs of AGN-type LINERS which are similar to UGC FR I AGNs,
but tend to underpredict the optical and radio core flux in UGC FR I
galaxies. For the well-studied galaxy M87, jet viewing angles in the
range $\sim 20\deg - 40\deg$ have been reported (Biretta, Zhou \& Owen
1995; Biretta \etal 1999). Already for $\gamma = 2$, logarithmic
Doppler boosting factors ${\cal B}$ (cf.~\ref{s:boosting}) vary by
$\Delta{\cal B}=0.91$ and $\Delta{\cal B}=-1.5$ between viewing angles
$\theta=30\deg$ and $\theta=90\deg$ for $p=2$ and $p=3$,
respectively. Thus the factors by which the optical and radio core
emission is underpredicted in the ADAF model for M87 could be on the
order of the variation in Doppler deboosting factors with jet viewing
angle.

M87, the nearest galaxy in the sample, offers the possiblity to
directly localize the origin of the radio core emission at the highest
attainable spatial resolution. Junor, Biretta \& Livio (1999) obtained
extreme high-resolution VLBI imaging (beam size 0.33mas x 0.12mas) at
43 GHz of the core of M87. The image shows that the jet does not
obtain its final collimation (as observed at pc and kpc scale) until a
few parsec from the core. Even at this high spatial resolution, there
is still a dominant core component of size $\lta 0.5$mas. This
corresponds to 0.04 pc or 130 Schwarzschild radii ($r_s=2GM_{\rm
BH}/c^2$) for M87. The peak flux density is 228 mJy/beam, while the
VLBA core flux density at 1.49 GHz is 1570 mJy/beam at a resolution of
10mas x 10mas. Hence a decrease by a factor of 2500 in beam area
corresponds to only a decrease by a factor 7 in peak flux density. The
VLBA core emission could therefore originate at the same scale as the
VLBI core emission. The flux difference might then be due to radio
core variability. Alternatively, the flux difference between 1.4 GHz
and 43 GHz corresponds to a radio spectral index of 0.57. Thus, it is possible for at
least M87 that the VLBA core flux (FWHM $\sim
0.01''$) is dominated by a nuclear component
which resides at radii smaller than those where jet collimation takes
place.

Optical spectral indices for an ADAF can vary considerably as a
function of mass accretion rate (Narayan, Mahadevan \& Quataert
1998). In contrast, the radio spectral index between $\sim 1- 100$ GHz
is always inverted with $\alpha \sim -0.4$ (Yi \& Boughn 1998)
compared to $\alpha \gta 0$ typically for jet
models. Table~\ref{t:spectralindex} lists the radio spectral index between
the 1.49 GHz / 1.67 GHz and 5 GHz core fluxes. Core flux variability
would have a significant effect on the derived spectral index due to
the small frequency range: a factor of two change in the flux density
ratio would result in a change by 0.6 in the derived spectral
index. For the spectral index between 1.49 GHz and 5 GHz, for which
the observations have similar resolution, 7 out of 10 sources have
spectra with $\alpha > -0.4$. For the spectral index between 1.67 GHz
and 5 GHz only 2 out of 10 have $\alpha > -0.4$. However, in the
latter case $\alpha$ is a lower limit as the 5 GHz flux is measured at
lower resolution than the 1.67 GHz VLBA measurements.

We conclude that at the {\sl arcsecond} scale the radio cores are most
likely dominated by the typical flat spectrum of a jet component
instead of a second non-relativistic component. However, it is unclear
at this point whether jet synchrotron or a second component, such as
ADAF synchrotron and/or inverse Compton emission, dominates the
$0.1''$ scale optical and $0.01''$ scale radio core
emission in all UGC FR I galaxies. Addressing this issue will require
knowledge of the core SEDs from radio to X-ray frequencies,
constraints on jet viewing angles and accretion flow and jet modeling.

\section{{\HalphaNII} emission excitation mechanism}
\label{s:halphanii}

Both the compactness of the core {\HalphaNII} emission and its strong
correlation with the radio and optical core emission from the AGN
indicate that the emission gas is excited by an AGN-related
process. The existence of young nuclear star clusters, which could
also produce the compact {\HalphaNII} emission, appears unlikely
(cf.~Section~\ref{s:cluster}). Excitation by old stars from the host
galaxy appears implausible as well, because the core {\HalphaNII}
luminosity does not correlate with either host magnitude or central
stellar surface brightness (as measured just outside the optical core;
cf.~Table~\ref{t:nuclum}). Two viable AGN-related excitation
mechanisms are photo- and shock-ionization. On the one hand the gas is
very close to the AGN which is generally believed to produce a
significant amount of photo-ionizing photons. On the other hand jets
are present, which might shock the central gas. For extended gas at
kiloparsec scales in nearby radio-loud galaxies (both FR I and FR IIs)
both excitation mechanisms often provide good fits to the observed
optical emission-line ratios (e.g., Baum, Heckman \& van Breugel 1992;
Simpson
\& Ward 1996). This degeneracy can be lifted by observing UV
emission-lines, for which the relative intensities are significantly
larger in the case of shock-ionization compared to photo-ionization
(Allen, Dopita \& Tsvetanov 1998).  This approach shows for M87 that
the gas as close as 50 pc from the nucleus is almost certainly
shock-excited (Dopita \etal 1997). As a reference, the core
{\HalphaNII} flux in M87 contributes about 20\% to the total flux
inside an aperture with a 50 pc radius and this radius is typically
about the size of the FWHM ($\sim 0.12''$) of our {\HalphaNII}
cores. M87 is a special case in the sense that it sits at the center
of a cooling flow which might cause enhancement of shock excitation.
Photons ionizing the gas can be produced by the accretion disk and
jet.  However, any photo-ionizing flux from the jet is highly beamed
even for the relatively low Lorentz factors inferred in
Section~\ref{s:boosting}.  The typical opening angle due to boosting ,
$1 / \gamma$ (e.g., Rybicki \& Lightman 1979) is $30\deg$ and $10\deg$
for $\gamma=2$ and $\gamma=5$ respectively.  Photo-ionization by the
accretion disk appears more likely.

Thus AGN processes certainly excite the central gas in the UGC FR I
galaxies but it cannot be determined at this point whether shock- or
photo-ionization is the dominant mechanism. We have recently obtained
HST/STIS optical spectra which can shed more light on this issue
(Noel-Storr \etal 2000).

\section{Connections between FR Is and low-luminosity AGNs}
\label{s:fr1sandllagns}

\subsection{AGN-type LINERS}
How similar are the central engines of UGC FR I nuclei and AGN-type
LINER nuclei which have radio cores but no large-scale radio jets? The
radio and optical core emission properties of the three AGN-type
LINERS discussed in Section~\ref{s:cluster} roughly follow the core
emission correlations found for the UGC FR I sample galaxies. For this
reason we construct a larger sample of nearby early-type galaxies with
LINER type nuclei and radio core emission. Ho \etal ( 1995,1997)
identified LINER nuclei among a sample of 486 nearby galaxies in the
RSA catalogue (RSA; Sandage \& Tammann 1981). Many of these LINER
nuclei were surveyed for radio core emission by Wrobel \& Heeschen
(1991). Ho (1999a) concluded that the radio core emission in these
LINER nuclei with radio core emission is non-thermal and most likely
produced by an AGN. In support of this interpretation, Falcke
\etal (2000) inferred brightness temperatures $T_B \gta 10^8$K from
higher resolution radio observations for a few AGN-type
LINERS. Further support for an AGN interpretation comes from the fact
that the five UGC FR I nuclei in the Ho~\etal sample (NGC 315, NGC
4261, NGC 4374, NGC 4486 and NGC 7626) are also classified as LINER
nuclei. The UGC FR I host galaxies are classified as E and S0 galaxies
(see paper I). The morphology classification in paper I and in Ho
\etal (1995) is not always consistent: NGC 315 is classified as an E
in paper I but as a S0 in Ho
\etal and vice versa for NGC 4374. We therefore selected all non-dwarf
galaxies in the Ho \etal sample (and not in the UGC FR I sample) which (
i) have morphologies which include `E' or `S0' in their description,
(ii) have an optical spectral classification which includes `LINER' in
their description and (iii) were surveyed for radio emission by Wrobel
\& Heeschen (1991). The emission-line flux for this `LINER sample' of 23
galaxies was measured through an aperture centered on the nucleus
which was typically $2'' \times 4''$ (Ho \etal 1997). The VLA
measurements at 5 GHz by Wrobel \& Heeschen (1991) have a FWHM of
$5''$. For 16 galaxies HST/ WFPC2 $V$- and/or $I$-band imaging is
available. We determined the NOS fluxes and upper limits in these
galaxies applying the same procedure as used for the UGC FR I sample
(see Section~\ref{s:opticalcores}). Only two LINER sample galaxies
have a NOS detection. Radio and optical core and central {\HalphaNII}
fluxes for the LINER sample are listed in Table~\ref{t:nucfluxliners}.
The errors in these core fluxes are similar to those of the UGC FR I
sample.

Figure~\ref{f:nucfluxliners}, which is similar to
Figure~\ref{f:nucflux}, shows the core emissions for both the UGC FR I
and LINER sample galaxies. For the UGC FR I sample
Figure~\ref{f:nucfluxliners} plots the {\HalphaNII} emissions inside a
radius of $1''$ (paper I) instead of the unresolved core {\HalphaNII}
emissions of Figure~\ref{f:nucflux} to provide a better match to the
apertures used for the LINER sample. All but five of the LINER sample
galaxies have radio core luminosities below those of the UGC FR I
sample. These five galaxies are discussed in more detail below. The
{\HalphaNII} luminosities of the two samples overlap, but the average
{\HalphaNII} luminosity is lower for the LINER sample. It is not clear
if the NOS luminosities for the two samples overlap as 14 of the 16
measurements are upper limits.

The core radio flux density and central {\HalphaNII} flux in the LINER
sample nuclei do not show the tight correlation observed in the UGC FR
I nuclei. At similar core radio flux densities, the {\HalphaNII}
fluxes of the LINER sample are typically more than 1 order of
magnitude larger than those of the UGC FR I sample. An offset remains
when one takes the radio core emission for the UGC FR I sample at
different resolutions (i.e., the 1.4 GHz or 5 GHz VLA
observations). The cores of the UGC FR I sample are expected to have
flat radio SEDs as function of $\nu$ so frequency differences are
not expected to alter the observed offset significantly. In the
luminosity-luminosity plane the LINER sample seems less offset, but
this could be due to the fact that both quantities depend on distance.
The offset is smaller for the elliptical LINER galaxies than
later-type LINER galaxies. In fact, the elliptical LINER galaxies are
consistent with following the luminosity correlation of the UGC FR I
sample. This was also noted by Ho (1999a) who argued that a correlation
is present even after taking into account the common dependence on
distance and host magnitude. The {\HalphaNII} emission for the two
samples is integrated over similar apertures, while the core emission
for the UGC FR I sample is observed at much smaller resolution than
the core emission for the LINER sample. The offset therefore could
suggest that the UGC FR I and elliptical LINER galaxies have an
additional source of radio emission compared to the later-type LINER
sample galaxies. The jets, clearly present in UGC FR Is at kpc scales,
are a plausible source for the extra radio core emission in at least
the UGC FR I galaxies. Alternatively, the enhanced emission-line
luminosity in later-type galaxies might reflect a larger amount of
interstellar matter present in these galaxies. Core radio and
emission-line observations for LINER galaxies at similar spatial
resolutions as the UGC FR I sample observations are needed to confirm
the observed trends.
 
Figure~\ref{f:nucfluxliners} plots $I$-band optical core emission for
all UGC FR I and most of the LINER sample. For a few LINER sample galaxies
$V$- or $R$-band measurements are used. The $V$- and $I$-band optical
core measurements generally differ by less than 50\%
(cf.~Section~\ref{s:opticalcores}) for the UGC FR I sample. This
difference is well within the scatter of the correlations. The $V$-
and $I$-band measurements differ by less than 50\% as well for NGC
4278, which is the only LINER sample galaxy with a NOS detection in
both bands. It is not possible to draw any firm conclusions from the
two correlations which involve the optical core flux measurements,
because most optical core flux measurements are upper limits. The
LINER sample is only consistent with following the two core emission
correlations as observed for the UGC FR I sample.

There are five LINER sample galaxies that have radio core luminosities
comparable to those observed in the UGC FR I sample: NGC 4278, NGC 4589,
NGC 5322, NGC 5353, and NGC 5354. Both NGC 5353 and NGC 5354 have a
compact radio source unresolved at sub-arcsecond resolution (Wrobel
1991; Filho, Barthel \& Ho 2000). NGC 4589 is unresolved at $5''$
resolution (Wrobel \& Heeschen 1991). Both NGC 4278 and NGC 5322 have a
jet-like extended radio component (Wrobel \& Heeschen 1984; Feretti
\etal 1984). In this respect it is also interesting that there is
evidence for a jet like structure in NGC 6500 (Falcke \etal 2000) and
NGC 3031 (Falcke \& Biermann 1999), which are two of the LINERS
discussed in Section~\ref{s:cluster}. HST/WFPC2 imaging is available
for NGC 5322, NGC 4589 and NGC4278. An apparently edge-on dust disk
obscures the galaxy nucleus in NGC 5322 and inhibits a NOS flux
measurement. The NOS detection and upper limit in NGC 4278 and NGC
4589 respectively are consistent with being on the radio - optical
core and {\HalphaNII} - optical core correlation (both in flux and
luminosity) observed for the UGC FR I sample. It will be worthwile to
obtain narrow-band and spectroscopic observations for the AGN-type
LINERs at higher spatial resolutions as performed for the UGC FR I
sample (cf.~Section~\ref{s:halphanii}). Such observations should
clarify to which extent the optical continuum flux is contaminated by
line-emission. For the optical continuum emission in UGC FR I nuclei
the contribution from line emission is small
(cf.~Section\ref{s:opticalcores}) but this is not necessarily the case
for the weaker cores in AGN-type LINERs (Cappellari
\etal 1999).

The conclusion is that AGN-type LINERS as a class do not seem to
follow the same core emission correlations as the UGC FR I sample in
general. The core emissions in LINER ellipticals however might lie on
the same correlations, but higher resolution radio and emission-line
observations are needed to confirm this. If true, this and the offset
in the distribution of radio core and central {\HalphaNII} fluxes
between the UGC FR I and elliptical LINERS on the one hand, and
later-type AGN-type LINERS on the other hand and the detection of
small-scale jets in some of the LINER galaxies then suggest: (i) the
central engines in FR I galaxies and AGN-type LINERS with elliptical
hosts are similar but the latter class are a down-scaled version of
the former, and (ii) the core emission, both at VLA and VLBA scales,
is dominated by inner jet emission.

\subsection{Kinematically Decoupled Cores}
One of the two NOS detections in the LINER sampole, NGC 4278, has a
kinematically distinct stellar core (KDC). This galaxy is part of a
HST/WFPC2 study of 18 elliptical galaxies with a KDC by Carollo \etal
(1997a,b).  They found an unresolved optical core in four galaxies
with $V-I$ colors similar to the UGC FR I optical cores. Analysis of
the nuclear emission in NGC 4278 and NGC 4552 suggests that both host
a low-luminosity AGN (Ho, Fillipenko \& Sargent 1997; Cappellari \etal
1999). Thus it is reasonable to assume that the optical core emission
in the other two KDC galaxies is produced by a low-luminosity AGN as
well. In fact, the KDC galaxies also have radio core and central
{\HalphaNII} emission (cf.~Table~\ref{t:nucfluxliners}). The squares
in Figure~\ref{f:nucfluxliners} show that also the radio and optical
core and central {\HalphaNII} emissions of KDCs roughly follow the
correlations observed for the UGC FR I cores.  The $\sim 20\%$ optical
core detection rate in KDCs at luminosities comparable to those
observed in FR I nuclei suggest that perhaps the peculiar central
dynamics of KDCs play a role in triggering nuclear activity.
Interestingly, one UGC FR I galaxy, NGC 7626, hosts a KDC as well.

\section{Summary \& Conclusions}
\label{s:summary}

In this paper we have analyzed the
relation between $0.01''$ scale radio and $0.1''$ scale optical
continuum and {\HalphaNII} core emission of a complete sample of 21
nearby FR I galaxies. The main conclusions are:
\begin{enumerate}
\item{We confirm the linear correlation between optical and radio core
emission in nearby FR I nuclei.  We find that both core emissions
also correlate with the core {\HalphaNII} emission. The mutual
correlations are unlikely to be caused by obscuration from the
ubiquitous central dust.}
\item{
Nuclear stellar clusters are highly unlikely to be the source of the
optical core emission for two reasons. First, previous spectral
studies have directly excluded a nuclear stellar cluster for two
typical members of the UGC FR I sample. Second, the UGC FR I radio,
optical and {\HalphaNII} core luminosities resemble the nuclei of
AGN-type LINERS more closely than stellar-type LINERS. An AGN origin
for the optical cores in all UGC FR I galaxies is strongly suggested
by the tight correlation with radio core emission, which is certainly
produced by the AGN.}
\item{
A jet origin for both the radio and optical core emission is favored
because (i) optical and radio core emission are tightly correlated,
(ii) spectral indices from radio to optical are similar to those for
extended optical jets, (iii) there is a suggestive trend with
independent estimates from jet orientation, and (iv) the correlation
residuals from both core emission with {\HalphaNII} emission are
corrrelated. However, a significant contribution from a second
component, such as accretion disk/flow/wind, to the radio and optical
emission at the $0.1''$ scale for the optical and the $0.01''$ scale
for the radio emission might be present.}
\item{The correlation of the optical and radio core emission with the
isotropic {\HalphaNII} emission constrains the core bulk Lorentz
factor $\gamma \lta 2$ if the inner jets consist of discrete blobs
($p=3$). For a continuous jet, bulk Lorentz factors $\gamma \sim 2-5$
are inferred. This result crtically depends on the assumed range in
viewing angle, which is assumed to be $[30\deg,90\deg]$. The bulk
Lorentz factors required by the BL Lac - FR I unification scheme are
generally larger, i.e..~$\gamma > 5$. If the core emission is
dominated by a jet component, this discrepancy could be reconciled by
a two-layer jet with a fast moving spine surrounded by a slower outer
layer.}
\item{The central gas is excited by AGN-related processes. Both shock- and
photo-ionization appear plausible excitation mechanisms at this time.}
\item{Radio, optical and {\HalphaNII} core luminosities
of elliptical LINER-type AGNs with and without kiloparsec-scale radio
jets appear to have similar relations. The results suggest (i) the
engines in the two types producing the cores might be similar and (ii)
the core radio emission is dominated by inner jet emission.}
\end{enumerate}

Conclusions 1 to 5 prompt us to make the following speculation: if the
radio and optical core emission are indeed inner jet synchrotron
emission, then their strong correlation with {\HalphaNII} core
emission implies either a direct link between jet luminosity and gas
excitation power, possibly via jet - gas interactions or a close
relation between AGN photo-ionizing power and jet radiative (and
possibly kinetic) power.

Conclusion 6 and the close resemblance of quiescent and radio-loud
active galaxies from scales just outside the AGN to the entire galaxy
and even environment lead to two further speculations. First, either
all bright early-type galaxies can host an AGN, or the capability of a
bright early-type galaxy to become an AGN is set by conditions within
the central few parsecs, currently unresolved at most
wavelengths. Second, the capability of a LINER-type AGN to become
radio-loud is set by conditions at similar scales. Two main
ingredients for an active nucleus are present at these scales:
accreting matter and a supermassive black hole . Likely factors
determining the formation of large-scale radio jets are then the black
hole spin and/or the inner accretion disk properties. A rigorous
comparison between the nuclei of active and quiescent nearby galaxies
at the tens of parsec resolution could advance our understanding on
these issues.


\acknowledgments

Support for this work was provided by NASA through grant number
\#GO-06673.01-95A from the Space Telescope Science Institute, which is
operated by AURA, Inc., under NASA contract NAS5-26555.  It is a
pleasure to thank Michele Cappellari for careful reading of the
manuscript and useful suggestions and the referee for helpful
suggestions that improved the manuscript. We thank Aaron Barth for
pointing out a significant typo in an earlier version of the paper.


\clearpage

\appendix
\section{Isophotal analysis of UGC 7115 and 3C 449}
\label{a:isophotes}

We present observations and the isophotal analysis for UGC 7115 and 3C
449 (UGC 12064). Together with the 19 galaxies presented in paper I
these galaxies form a well-defined complete sample of FR I galaxies.
For UGC 7115 we present new HST/WFPC2 $V$- and $I$-band imaging, while
for 3C 449 we are using archival WFPC2 $R$-band and narrow-band
imaging presented in Martel
\etal (1999) and Martel \etal (2000) (see Figure~\ref{f:imv}). 
The data reduction for these two galaxies and the rest of the sample
was done in similar fashion and is discussed in detail in paper
I. Some general properties for the two galaxies are listed in
Table~\ref{t:generalproperties} and the observation logs are presented
in Table~\ref{t:obs}.

\subsection{UGC 7115}
UGC 7115 is a relatively isolated galaxy. The nearest galaxy at
similar distance as catalogued by NED\footnote{The NASA/IPAC
Extragalactic Database (NED) is operated by the Jet Propulsion
Laboratory, California Institute of Technology, under contract with
the National Aeronautics and Space Administration.} is at
$9.5'$. Figure~\ref{f:imv} shows the WFPC2 $V$-band image of UGC 7115.
We detect an almost round dust disk (axis ratio $\sim 0.95$) with a
diameter of $1.3''$ (570 pc). The rim of the disk appears darker than
its inner region. UGC 7115 has a central blue nuclear optical source
(see Section~\ref{s:opticalcores}). We compare our isophotal results
(see Figure~\ref{f:iso}) with isophotal results for $r=2''-20''$ from
cousins $R$-band imaging of UGC 7115 by Fasano
\& Bonoli (1989) taken under bad seeing. The predicted $R$
magnitude from the $V$ and $I$ band imaging is $\sim$ 1 magnitude
brighter than observed by Fasano \& Bonoli. Moreover, they derive an
isophotal PA $\sim 90\deg$ for $r=5''-10''$ which increases to
$110\deg$ inwards. The cause of the $\sim 1$ magnitude difference in
photometric calibration and the $\sim 90\deg$ offset in PA is
unknown. The latter difference might be due to the bad seeing. Lambas,
Groth \& Peebles (1988) report a PA=$170\deg$ consistent with our
results. The ellipticity derived by Fasano \& Bonoli agrees to within
$\sim 0.02$. UGC 7115 has a one-sided jet structure (Xu \etal 2000).

\subsection{3C 449 (UGC 12064)}
3C 449 has a companion at $\sim 37''$ and is part of Zwicky cluster
2231.2 +3732. The HST/WFPC2 $R$-band imaging for 3C 449 was presented
by Martel \etal (1999) and is shown in Figure~\ref{f:imv}. The galaxy
has a central dust disk $3.5''$ (1.2kpc) in diameter with a mottled
morphology and was detected by Capetti \etal (1994). The disk is aligned
with the stellar major axis. Our estimate of the disk inclination is
$57 \pm 3 \deg$, assuming the disk is intrinsically circular. This
inclination is a bit lower than the $\sim 65\deg$ reported by Martel
\etal (1999). The {\HalphaNII} emission is concentrated on the nucleus
and there is a spot at $1.3''$ North from nucleus on the inside of the
dust ring. We compare our isophotal analysis (see Figure~\ref{f:iso})
with De Juan, Colina \& P\'{e}rez-Fouron (1994), who used ground-based
observations. The ellipticity profile agrees well outside the dust
disk radius and $\epsilon$ increases towards 0.3 between
$r=10''-30''$. The PA agrees as well and continues to increase to
$240\deg$ at $r=30''$. 3C 449 has a well-known twin-jet radio
structure (Feretti \etal 1999, and references therein).

\section{Beaming parameters}
\label{a:beaming}
We want to estimate how Doppler boosting affects the observed relation
between two fluxes $f_1$ and $f_2$ for which the rest-frame fluxes ${f'}_1$ and 
${f'}_2$ are
related by a power law defined as
\begin{equation}
\log {f'}_2 = a_{\rm int} \log {f'}_1 + b_{\rm int}.
\end{equation}
The primed quantities are measured in the rest frame of the
emitter. The flux $f_1$ is not affected by beaming ($f_1={f'}_1$) but
the monochromatic flux $f_2$ is. 
The flux $f_{\nu}(\nu)$ and
${f'}_{\nu}({\nu'})$ for an emitting particle are related by (e.g., Urry \& 
Padovani, 1995):
\begin{equation}
f_{\nu}(\nu) = \delta^p {f'}_{\nu'}(\nu') ,
\end{equation}
where $\nu$ and $\nu'$ are related as $\nu= \delta \nu'$ and the
kinematic Doppler factor $\delta$ is defined as
\begin{equation}
\delta = [\gamma(1 - \beta \cos \theta)]^{-1} ,
\end{equation}
with $\gamma=\sqrt{1/(1-\beta^2)}$ the Lorentz factor and $\beta$ the
velocity of the emitting material in units of the speed of light, and
$\theta$ the angle of the velocity vector with the line of sight. The
factor $p$ depends on the structure of the jet and the spectral index
of the emitting material (cf.\ Urry \& Padovani 1995). For example,
for an isotropically emitting smooth continuous jet $p=2+\alpha$,
while for a moving discrete source with isotropic emission
$p=3+\alpha$, where $\alpha$ is the rest-frame spectral index defined
as ${f'}_{\nu'} \sim {(\nu')}^{-\alpha}$.
We consider the case where the flux $f_2$ is emitted by a symmetric two-sided Doppler boosted
jet and hence the relation between $f_2$ and ${f'}_2$ is
\begin{equation}
\log f_2= \log {f'}_2 + \log \frac{\delta^p(\theta)+\delta^p(\pi+\theta)}{2} 
\equiv \log {f'}_2 + {\cal B},
\end{equation}
where $\log$ indicates the logarithm of base 10. We will refer to
${\cal B}$ as the logarithmic Doppler boosting factor.

A linear least-squares regression fit to a sample of logarithmic flux
pairs $f_1,f_2$ will yield a slope $a_{\rm obs}$ and intercept $b_{\rm
obs}$ (in the absence of measurement errors):
\begin{equation}
\begin{array}{l}
a_{\rm obs} = a_{\rm int} + \frac{\overline{f_1{\cal B}} - \overline{f_1} \cdot 
\overline{\cal B}}{\overline{{f_1}^2} - {\overline{f_1}}^2}\\
b_{\rm obs} = b_{\rm int} + \overline{\cal B}, \\
\end{array}
\end{equation}
where the overlining denotes the mean of the quantity. We assume that
${\cal B}$ does not depend on either ${f'}_1$ or ${f'}_2$. In that case
we obtain $a_{\rm obs} = a_{\rm int}$. Thus the observed slope of the
linear regression fit for a sample of ($f_1$,$f_2$) pairs
will equal the slope of the intrinsic relation between ${f'}_1$ and
${f'}_2$ and  the observed residuals will equal ${\cal B} -
\overline{\cal B}$.
If we assume that the sample has spherically randomly distributed jet
viewing angles in the range $\theta=[\theta_0,\theta_1]$ and a known
identical bulk Lorentz factor $\gamma$ and jet structure parameter $p$
we can evaluate the mean and standard deviation of the logarithmic
Doppler boosting factor by
\begin{equation}
\begin{array}{ll}
\overline{\cal B} = \int_{\theta_0}^{\theta_1} {\cal B} \sin\theta d\theta / 
\int_{\theta_0}^{\theta_1} \sin\theta d\theta \\
\sigma^2({\cal B}) = \overline{{\cal B}^2}- {\overline{\cal B}}^2, \\
\end{array}
\end{equation}
The integral for $\overline{\cal B}$ can be expressed in terms of
elementary functions for integer $p$. The result for $p=1,2,3$:
\begin{equation}
\begin{array}{ll}
y \equiv \beta\cos\theta, \\
C \equiv \frac{\log e}{y_1 - y_0}, \\
I_0 \equiv [-p\ln(\gamma)y]^{y_1}_{y_0}, \\
I_1 \equiv -[(1+y)\ln(1+y)-(1+y)]^{y_1}_{y_0}, \\
I_2 \equiv [(1-y)\ln(1-y)-(1-y)]^{y_1}_{y_0}, \\
I_3 \equiv [y\ln(y^2+1) - 2y + 2\arctan(y)]^{y_1}_{y_0}, \\
I_4 \equiv \frac{1}{\sqrt{3}}[\sqrt{3}y\ln(3y^2+1) - 2\sqrt{3}y + 2\arctan(\sqrt{3}y)]^{y_1}_{y_0}, \\
\overline{\cal B}= C(I_0+I_1+I_2), & {\rm for} \quad p=1, \\ 
\overline{\cal B}= C(I_0+2[I_1+I_2]+I_3), & {\rm for} \quad p=2, \\ 
\overline{\cal B}= C(I_0+3[I_1+I_2]+I_4), & {\rm for} \quad p=3. \\ 
\end{array}
\end{equation}
The integral for $\overline{{\cal B}^2}$ must be evaluated
numerically. The intercept shift $\overline{\cal B}$ is not directly
observed, because $b_{\rm int}$ is unknown. However, the residual,
${\cal B}-{\overline{\cal B}}$, and the variance of the regression fit
$\overline{{\cal B}^2}$ are both observable. Thus one can determine
which combinations of $\gamma$ and $p$ are consistent with the
observed residuals of the regression fit and their variance.


\ifsubmode\else
\baselineskip=10pt
\fi


\clearpage


\ifsubmode\else
\baselineskip=14pt
\fi


\newcommand{\figcapnosexamples}
{Three examples of the central $V$-band flux profile (DN counts) along
the WFPC2/PC CCD line which crosses the galaxy nucleus in the WFPC2
image. NGC 383 (left) has an unresolved nuclear optical source (NOS)
which clearly stands out from the shallow-sloped stellar background
(the WFPC2/PC PSF has a FWHM of $\sim 2.2$ pixels). NGC 4335 (middle)
has a bright and steeply rising stellar flux profile which makes the
detection of the expected NOS flux very difficult. The central flux
profile in NGC 7052 (right) is severely affected by dust obscuration
which inhibits a reliable determination of the NOS flux. See
Section~\ref{s:opticalcores} for a complete discussion of the NOS
measurements.\label{f:nosexamples}}

\newcommand{\figcapnucflux}
{The VLBA radio core, WFPC2 optical core, and core {\HalphaNII}
emission in the UGC FR I nuclei plotted versus each other. The left
column contains flux-flux plots, the right column contains
luminosity-luminosity plots. Arrows indicate upper limits. The
correlation between each pair of quantities is significant at more
than the 99\% level according to a generalized Kendall's tau test (see Tables~\ref{t:nucflux} and
\ref{t:nuclum}). The error bars in the lower right corner of each plot
indicate the typical error for each quantity. The large arrow in the
lower right plot indicates the displacement of a datapoint if it were
observed through a dust screen with a $V$-band opacity
$A_V=3$\label{f:nucflux}}
 
\newcommand{\figcapnocradiocore} {Optical WFPC2 $I$-band core flux 
as a function of the radio core flux for the 3CR FR I sample (open
circles; Chiaberge, Capetti \& Celotti 1999) and for our UGC FR I
sample (filled circles; we plot our core emission measurements for the
sources overlapping with the 3CR sample). The crossed circle is NGC
6251 from Hardcastle \& Worrall (2000). Arrows indicate upper
limits. For the UGC sample the radio core flux at 1.49 GHz is
plotted, while for the other galaxies the 5 GHz core flux is plotted
(see Section~\ref{s:radionos} for discussion). The outlier is the
peculiar galaxy 3C 386 (see Section~\ref{s:radionos}). The combined
sample shows a highly significant linear correlation between radio and
optical core flux.\label{f:nocradiocore}}
 
\newcommand{\figcapresiduals}
{Both radio and optical core luminosity are well correlated with the
central {\HalphaNII} luminosity in log-log space
(Figure~\ref{f:nucflux}). This plot shows the residuals of the linear
regression fit to these two correlations. The linear regression fits
consider the {\HalphaNII} luminosity as the independent variable. The
arrows indicate the upper limits to the optical core emission. The two
residuals are again well correlated, especially when taking the
typical errors in the fluxes into account
(cf. Figure~\ref{f:nucflux}). This supports the idea that the radio
and optical core emission have a common origin, as discussed in
Sections~\ref{s:correlations} and
\ref{s:boosting}.\label{f:residuals}}
   
\newcommand{\figcapboosting}
{Doppler boosting of jet radio emission causes scatter around the
linear regression fit of the observed logarithmic radio flux as a
function of {\HalphaNII} flux (see Section~\ref{s:radiohalpha} and
Appendix~\ref{a:beaming}).  {\bf (a):} the predicted standard
deviation of this scatter $\sigma^2({\cal B})$ is plotted as a
function of the jet bulk Lorentz factor $\gamma$ for a continuous jet
($p=2$, thin curves) and a jet consisting of discrete blobs ($p=3$,
thick curves). For both models the rest-frame flux is emitted
isotropically and has a flat spectrum (i.e., $\alpha=0$). Jet viewing
angles are spherically randomly distributed in the range
$[20\deg,90\deg]$ (solid curves), $[30\deg,90\deg]$ (dotted curves)
and $[40\deg,90\deg]$ (dashed curves).

The observed standard deviation is $\sim 0.3$. Any inferred bulk
Lorentz factor is in fact an upper limit because there are probably
other sources of scatter such as measurement errors and flux
variability.  {\bf (b):} the predicted residual, ${\cal
B}-{\overline{\cal B}}$, from the correlation is plotted as a function
of jet viewing angle $\theta$ for a continuous jet (thin curves) and
a jet consisting of discrete blobs (thick curves). The jet parameters
are identical to those under (a). The three curves per jet type are
for $\gamma=2$, $\gamma=3$ and $\gamma=5$ (i.e., increasing maximum
and minimum residuals). The observed range of residuals is $\sim
[-0.5,1.0]$.\label{f:boosting}}

\newcommand{\figcapboostingc}
{Left: the residuals of the linear regression fits to the radio core
luminosity as a function of {\HalphaNII} core luminosity (in log-log
space) plotted as a function of radio core luminosity. Consistent with
a Doppler boosting origin for the residuals all galaxies with jets
expected to be oriented close to the plane of the sky (filled squares)
have negative residuals, while galaxies with jets expected to be
oriented close to the line of sight (filled circles) have close to
positive or positive residuals. These expectations are based on jet
properties (see Section~\ref{s:boosting}). Addtional support for the
beaming scenario is the trend that more luminous cores have more
positive residuals. Right: similar plot as the left panel but now for
the optical core emission. The plot is consistent with the conclusions
drawn for the radio cores. The increase in scatter in this plot is
expected from the larger measurement error in the optical core flux
compared to the radio core flux. Both trends are mainly
driven by the few points at the high and low end of the distributions
and hence larger samples are required to confirm these
results.\label{f:boostingc}}

\newcommand{\figcapliners}
{The correlations between the core luminosities for the UGC FR I
galaxies (dots) but now with the data for the UV-bright LINER sample
indicated as well (Maoz \etal 1998; cf.\
Table~\ref{t:liners}). Stellar-type LINER galaxies are denoted by
stars and AGN-type LINERS are denoted by triangles. Arrows denote
upperlimits. The AGN-type LINER and UGC FR I core emissions roughly
overlap, while the stellar-type LINERS deviate, mainly in their radio
luminosities. This suggests an AGN origin for all three fluxes in UGC
FR I nuclei (see Section~\ref{s:cluster}).
\label{f:liners}}
 
\newcommand{\figcapnucfluxliners}
{This figure is similar to Figure~\ref{f:nucflux} and plots the radio
core, optical core and central {\HalphaNII} emission of the UGC FR I
nuclei (solid dots), LINER sample nuclei (triangles) and ellipticals
with kinematically decoupled cores (KDCs, squares) discussed in
Section~\ref{s:fr1sandllagns}. The solid triangles are elliptical
galaxies and the open triangles are E/S0 and S0 galaxies (see
Table~\ref{t:nucfluxliners}). Arrows indicate upper limits and the
error bars in the lower right corner of each panel indicate the
typical errors. The plots suggest that the elliptical LINER and KDC
nuclei follow the core emission correlations observed for the UGC FR I
sample. See Section~\ref{s:fr1sandllagns} for further
discussion.\label{f:nucfluxliners}}
  
\newcommand{\figcapimv}
{$V$-band image of UGC 7115 and $R$-band image of 3C 449 (UGC
12064). The images have logarithmic stretch. North is up and East is
to the left.\label{f:imv}}

\newcommand{\figcapiso}{\footnotesize Isophotal parameters in the
$I$-band (UGC 7115) and $R$-band (3C 449) as a function of radius along
the major axis. From top to bottom: surface brightness, $V$-$I$ color,
position angle, ellipticity and fourth order Fourier coefficient. The
galaxy name is shown above each column. The top plot for UGC 7115 also
shows the $V$-band luminosity profile (i.e. higher $\mu$).  The dashed
and dotted lines display luminosity profiles corrected for dust
obscuration (see paper I). Error bars include the formal error given
by the isophotal fitting routine and the error in the estimated sky
counts. The vertical dashed line indicates the maximum radius out to
which the isophotes are affected by dust obscuration and PA and
$\epsilon$ are kept fixed.\label{f:iso}}


\ifsubmode 
\figcaption{\figcapnosexamples}
\figcaption{\figcapnucflux}
\figcaption{\figcapnocradiocore}
\figcaption{\figcapresiduals}
\figcaption{\figcapliners}
\figcaption{\figcapboosting}
\figcaption{\figcapboostingc}
\figcaption{\figcapnucfluxliners}
\figcaption{\figcapimv}
\figcaption{\figcapiso}
 \clearpage
\else\printfigtrue\fi

\ifprintfig

 
\clearpage
\begin{figure}
\epsscale{1.0}
\plotone{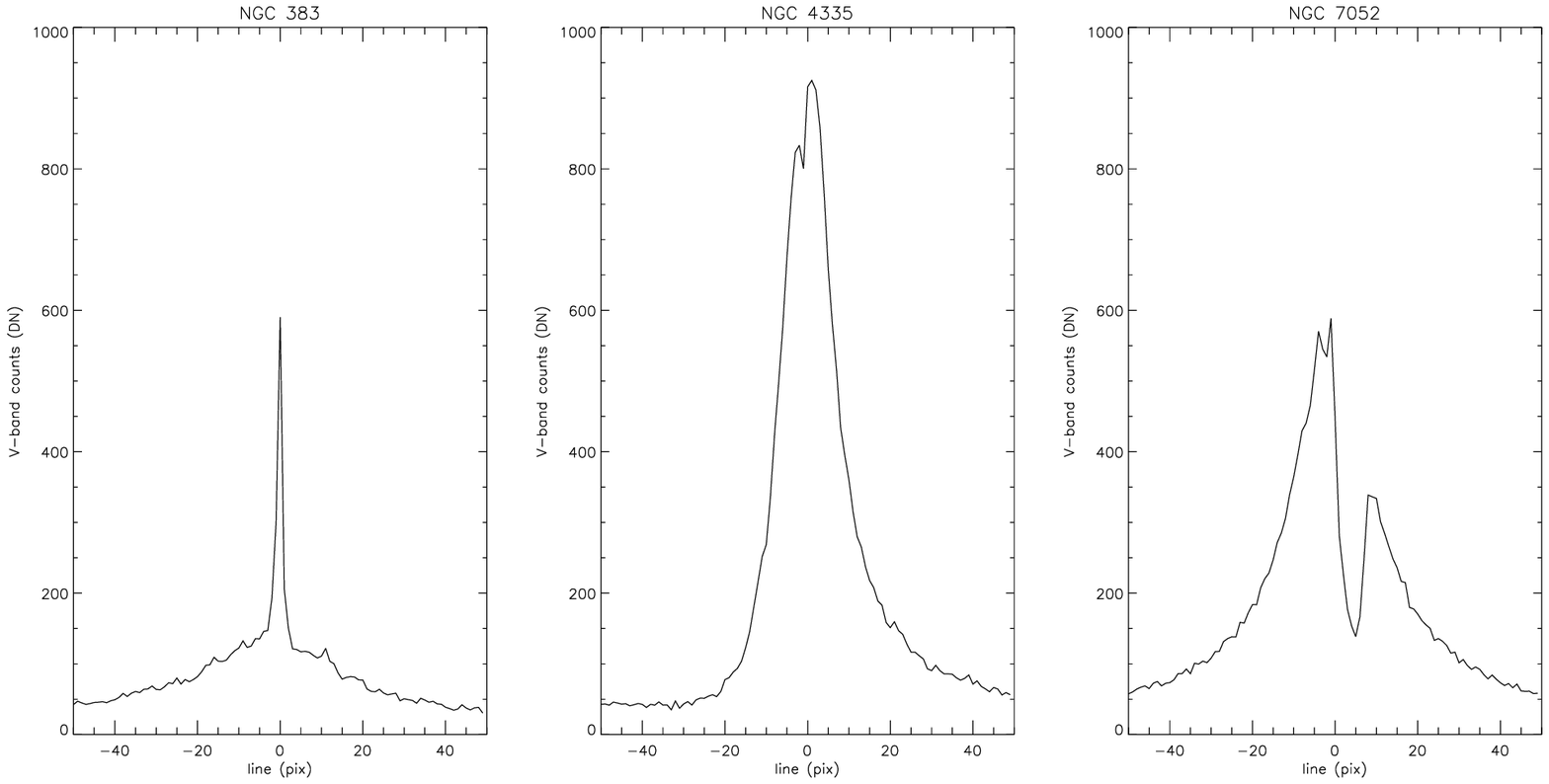}
\ifsubmode
\vskip3.0truecm
\setcounter{figure}{0}
\addtocounter{figure}{1}
\centerline{Figure~\thefigure}
\else\figcaption{\figcapnosexamples}\fi
\end{figure}

\clearpage
\begin{figure}
\epsscale{0.75}
\plotone{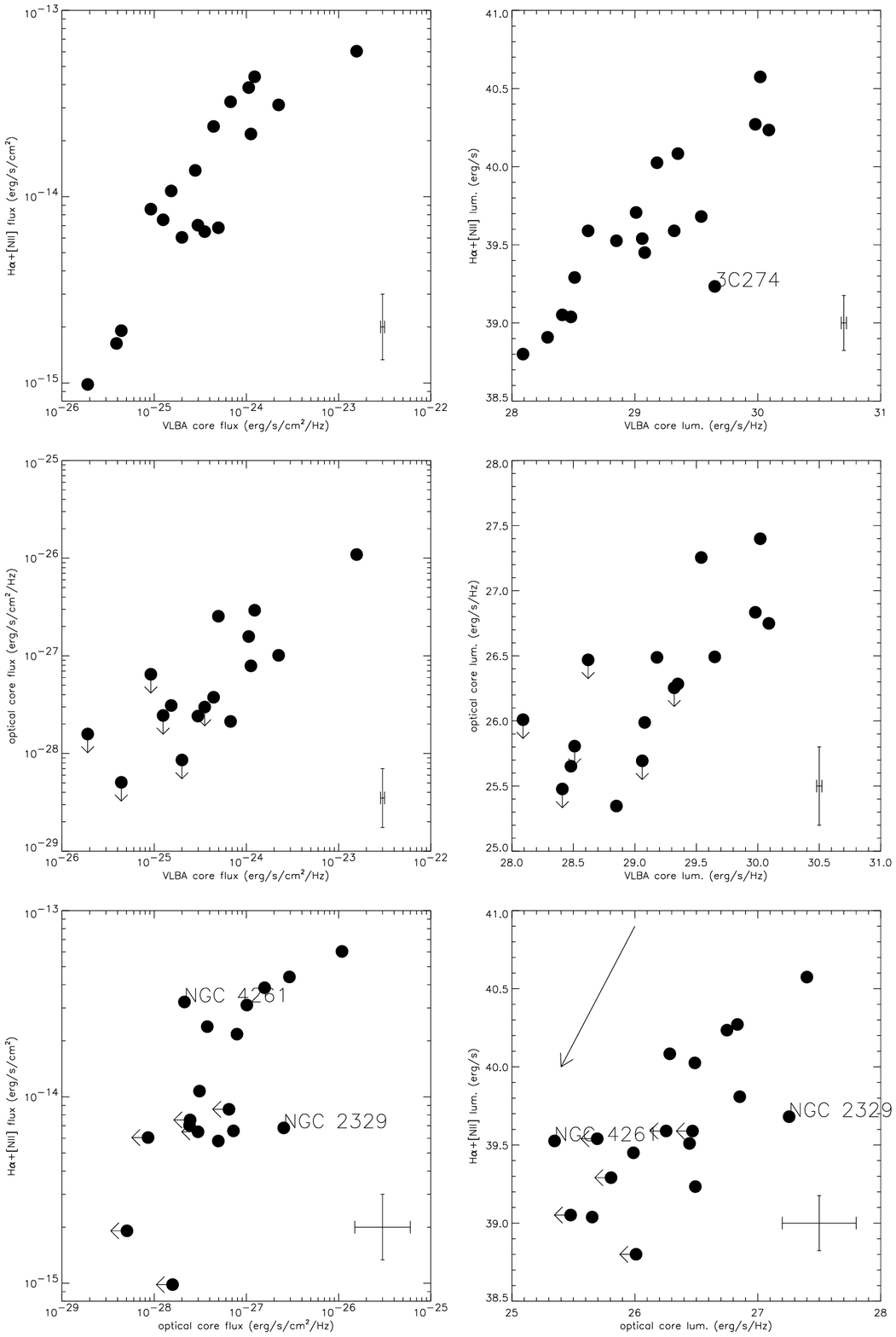}
\ifsubmode
\vskip3.0truecm
\addtocounter{figure}{1}
\centerline{Figure~\thefigure}
\else\figcaption{\figcapnucflux}\fi
\end{figure}

\clearpage
\begin{figure}
\epsscale{1.0}
\plotone{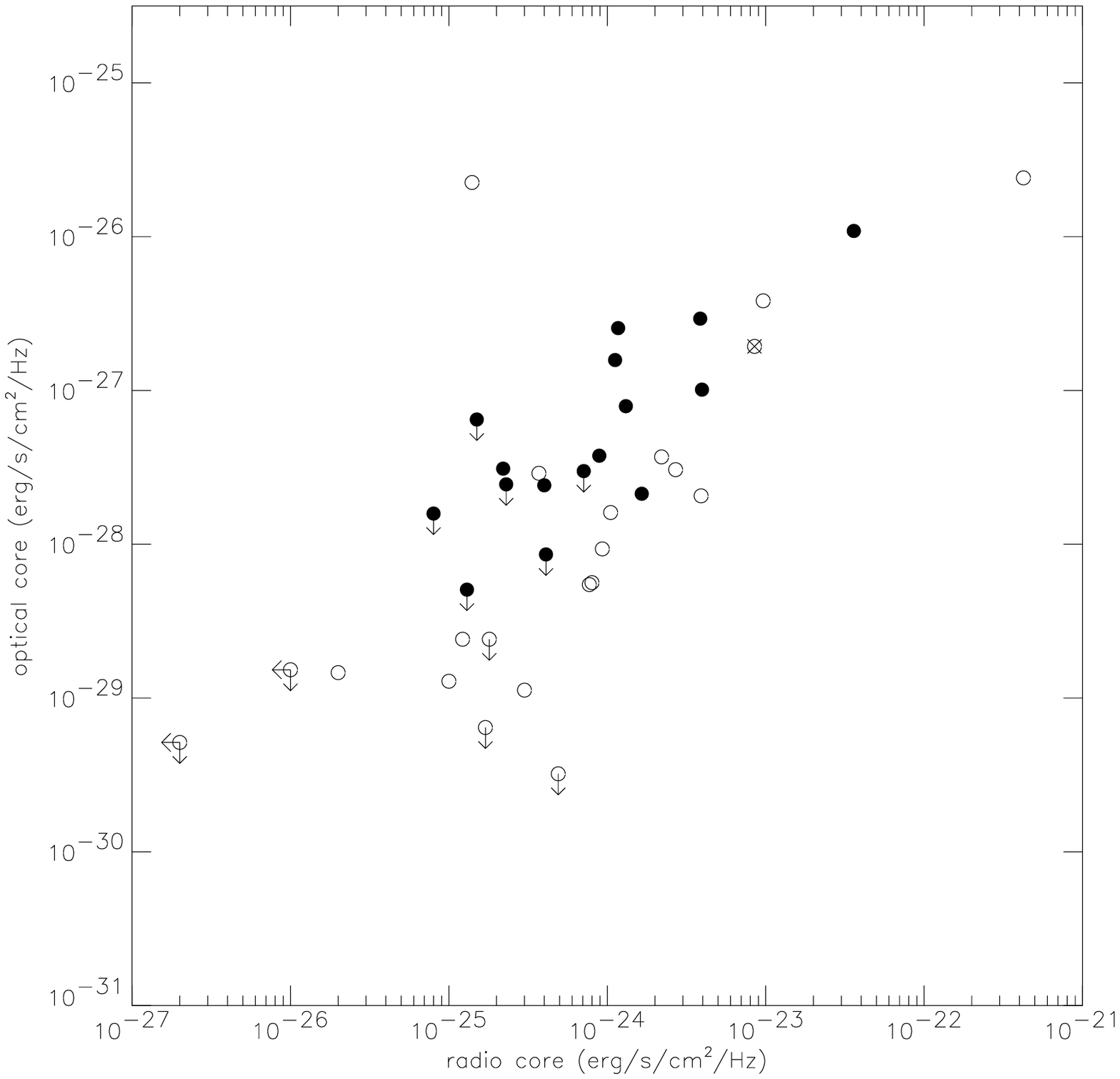}
\ifsubmode
\vskip3.0truecm
\addtocounter{figure}{1}
\centerline{Figure~\thefigure}
\else\figcaption{\figcapnocradiocore}\fi
\end{figure}

\clearpage
\begin{figure}
\epsscale{1.0}
\plotone{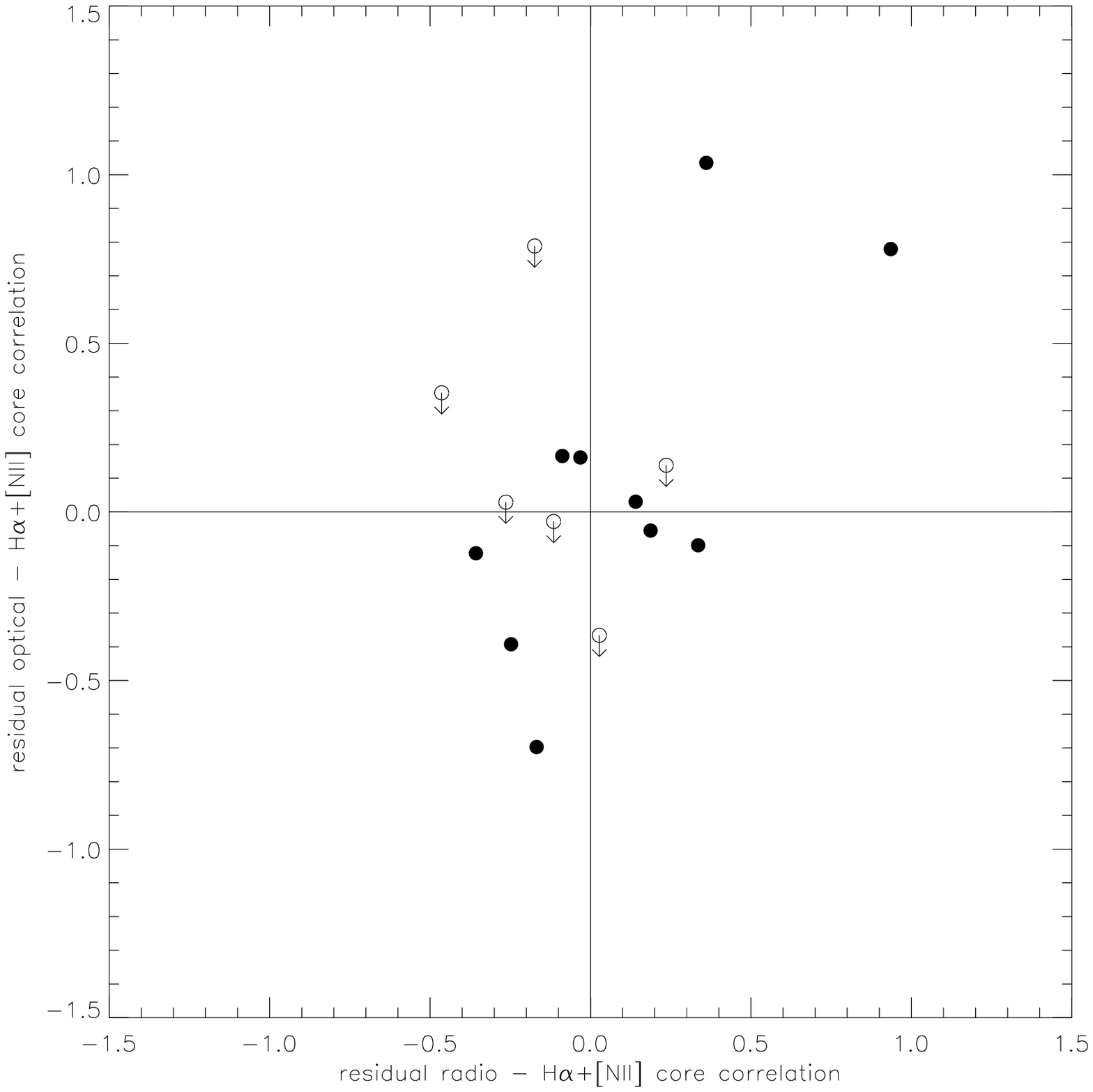}
\ifsubmode
\vskip3.0truecm
\addtocounter{figure}{1}
\centerline{Figure~\thefigure}
\else\figcaption{\figcapresiduals}\fi
\end{figure}

\clearpage
\begin{figure}
\epsscale{0.75}
\plotone{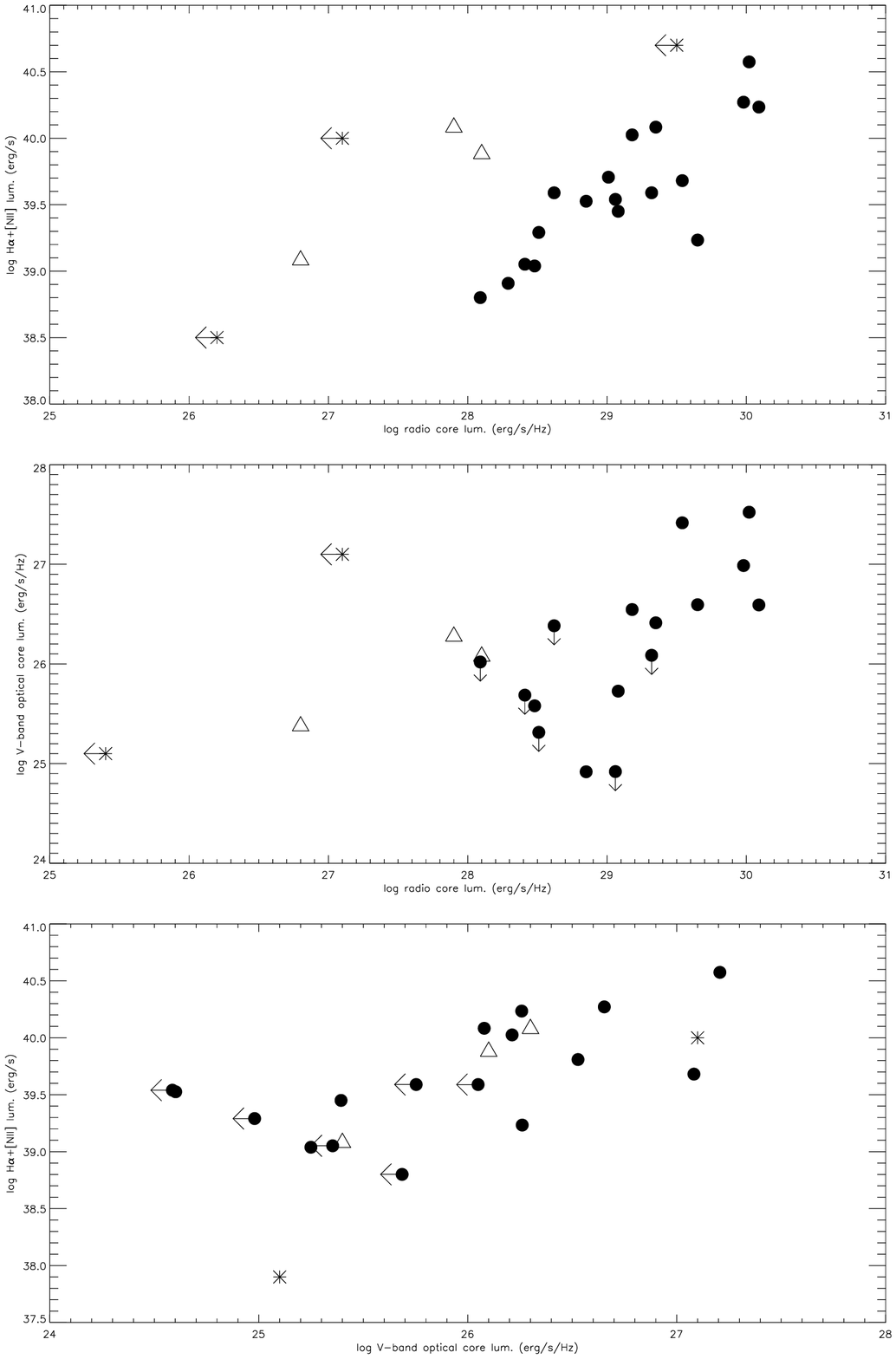}
\ifsubmode
\vskip3.0truecm
\addtocounter{figure}{1}
\centerline{Figure~\thefigure}
\else\figcaption{\figcapliners}\fi
\end{figure}

\clearpage
\begin{figure}
\epsscale{1.0}
\plotone{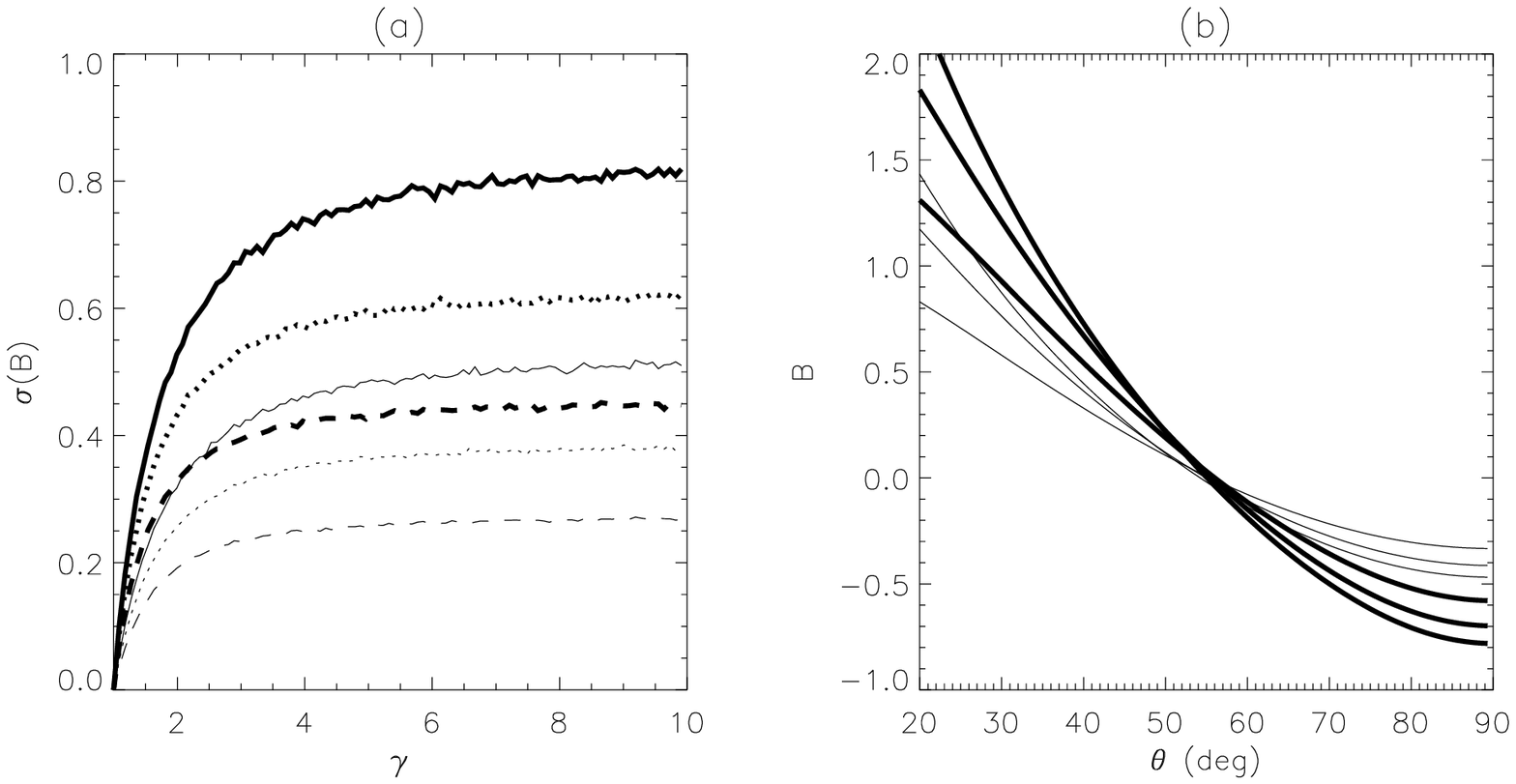}
\ifsubmode
\vskip3.0truecm
\addtocounter{figure}{1}
\centerline{Figure~\thefigure}
\else\figcaption{\figcapboosting}\fi
\end{figure}

\clearpage
\begin{figure}
\epsscale{1.0}
\plotone{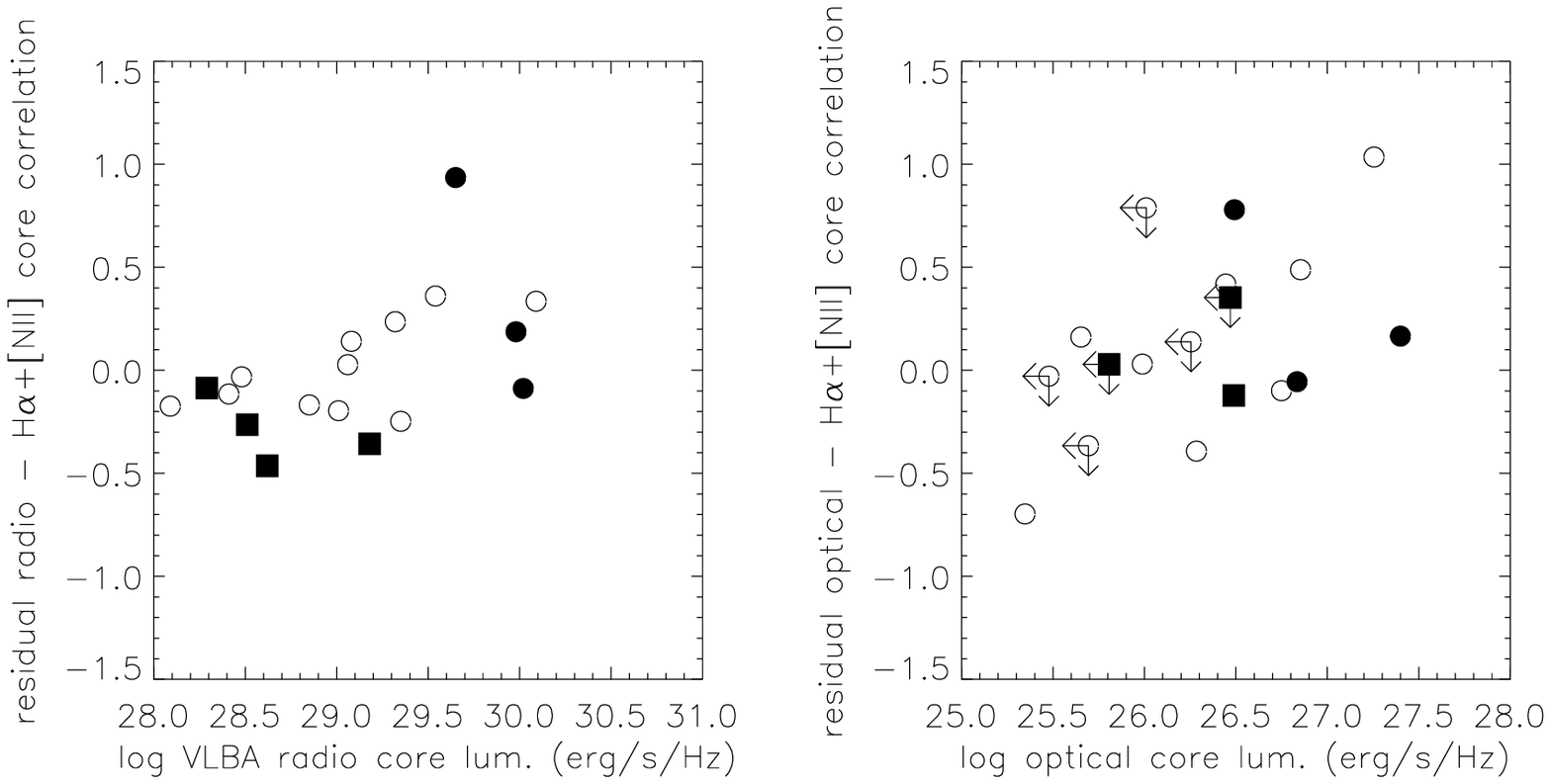}
\ifsubmode
\vskip3.0truecm
\addtocounter{figure}{1}
\centerline{Figure~\thefigure}
\else\figcaption{\figcapboostingc}\fi
\end{figure}

\clearpage
\begin{figure}
\epsscale{0.75}
\plotone{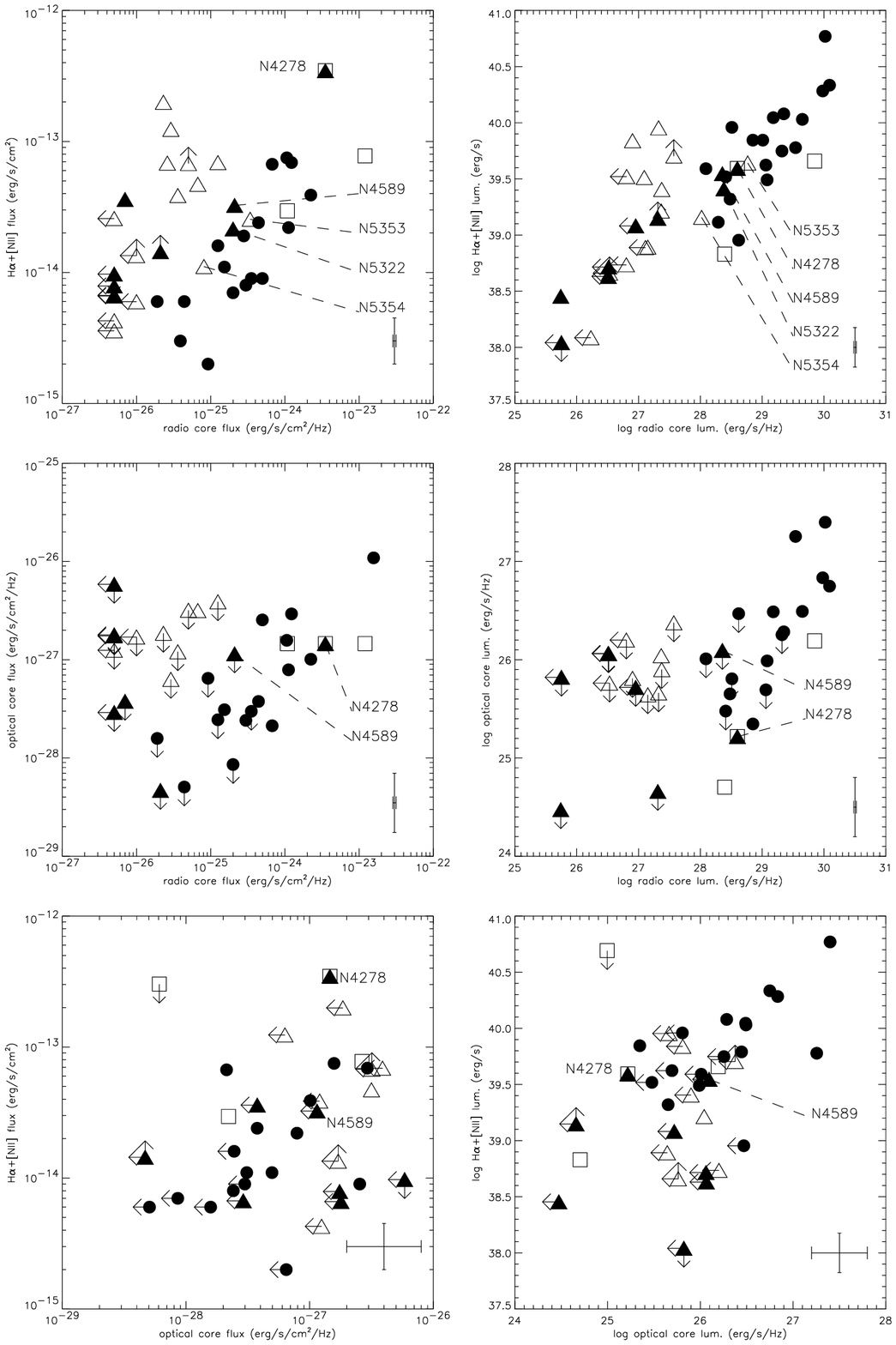}
\ifsubmode
\vskip3.0truecm
\addtocounter{figure}{1}
\centerline{Figure~\thefigure}
\else\figcaption{\figcapnucfluxliners}\fi
\end{figure}

\clearpage
\begin{figure}
\epsscale{1.0}
\plottwo{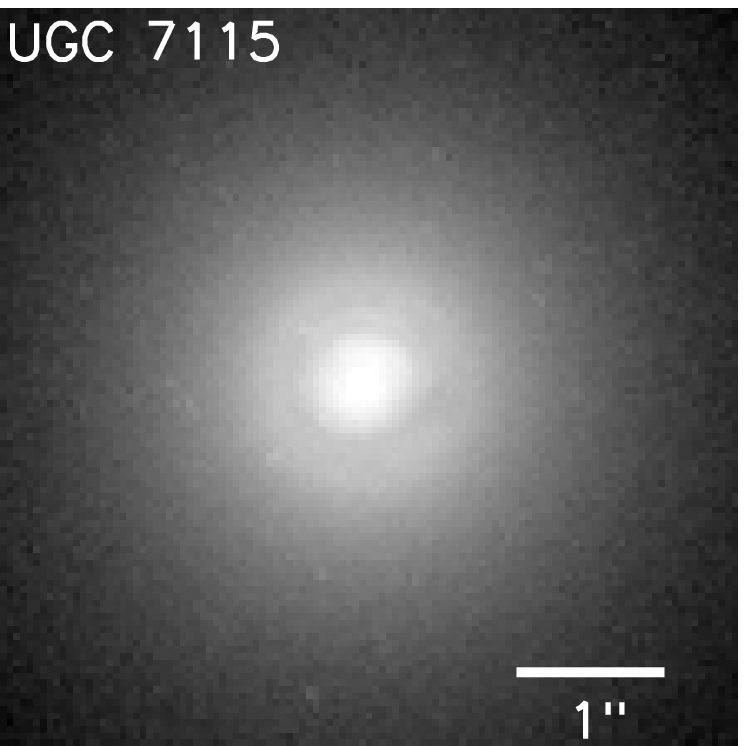}{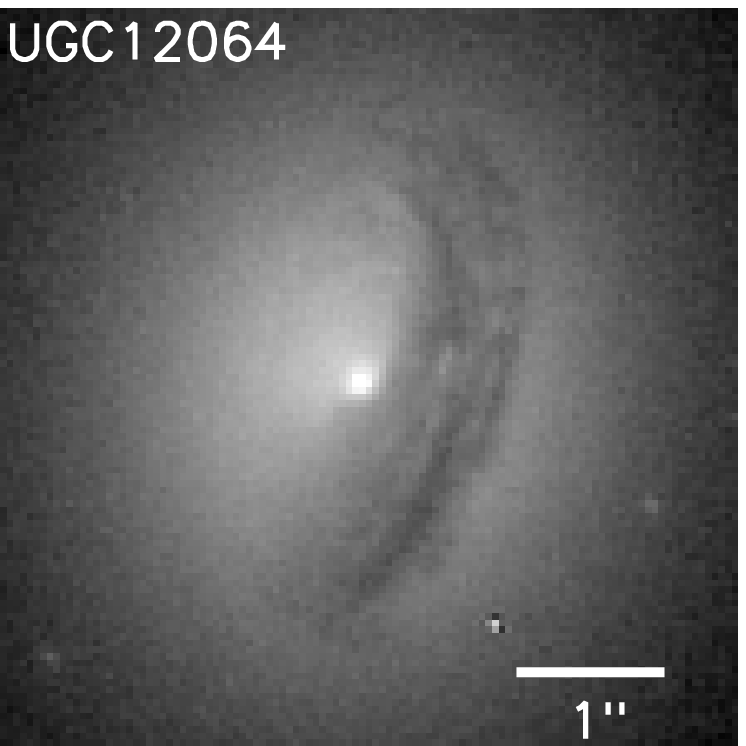}
\ifsubmode
\vskip3.0truecm
\addtocounter{figure}{1}
\centerline{Figure~\thefigure}
\else\figcaption{\figcapimv}\fi
\end{figure}
 
\clearpage
\begin{figure}
\epsscale{0.8}
\plotone{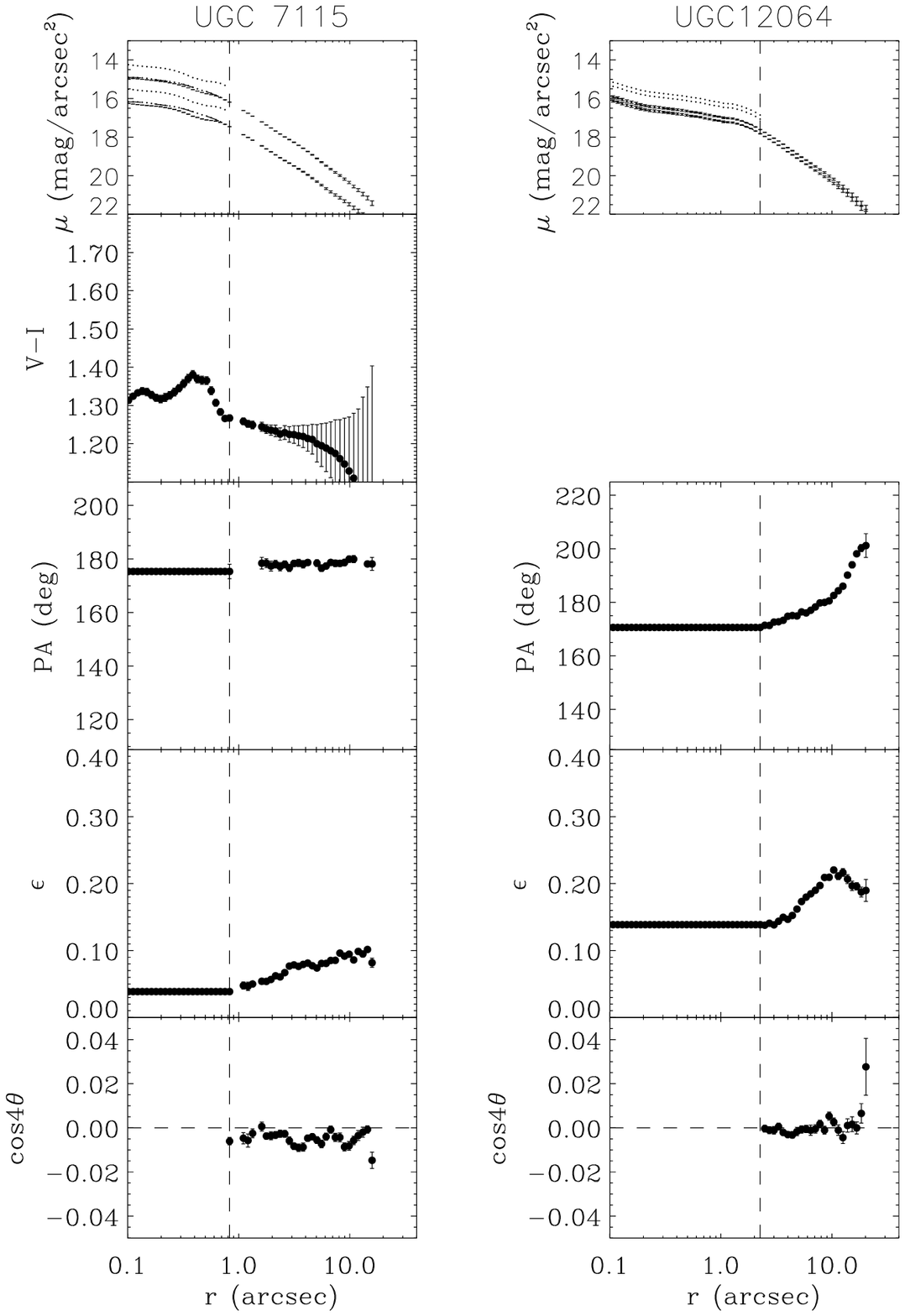}
\ifsubmode
\vskip3.0truecm
\addtocounter{figure}{1}
\centerline{Figure~\thefigure}
\else\figcaption{\figcapiso}\fi
\end{figure}
 

\fi


\clearpage
\ifsubmode\pagestyle{empty}\fi

\begin{deluxetable}{lllrlrr}
\tablewidth{0pt}
\tablecaption{HST/WFPC2 core fluxes\label{t:nucfluxdetection}}
\tablehead{
\colhead{Name 1} & \colhead{Name 2} & \colhead{Filter $V$} & \colhead{NOS flux 
$V$} 
& \colhead{Filter $I$} & \colhead{NOS flux $I$} & \colhead{{\HalphaNII} flux} \\
\colhead{(1)} & \colhead{(2)} & \colhead{(3)} & \colhead{(4)} & \colhead{(5)} & 
\colhead{(6)} & \colhead{(7)} \\
}
\startdata
NGC193  &         & F555W &  6.2e-18 & F814W &  1.1e-17 &      7.0e-15  \\
NGC315  &         & F555W &  3.3e-17 & F814W &  4.7e-17 &      3.1e-14  \\
NGC383  &         & F555W &  2.4e-17 & F814W &  1.8e-17 &      2.4e-14  \\
NGC541  &         & F555W & $<$7.5e-18 & F814W & $<$7.4e-18 &  9.8e-16 \\
NGC741  &         & F555W & $<$3.8e-18 & F814W & $<$2.4e-18 &  1.9e-15  \\
UGC1841 & 3C66B   & F555W &  5.2e-17 & F814W &  3.7e-17 &      2.2e-14  \\
NGC2329 &         & F555W &  1.7e-16 & F814W &  1.2e-16 &      6.8e-15  \\
NGC2892 &         & F555W &  1.6e-17 & F814W &  1.4e-17 &      1.1e-14  \\
NGC3801 &         & F555W & $^{\dag}$ & F814W & $^{\dag}$ & $<$3.3e-16 \\
NGC3862 & 3C264   & F547W &  1.9e-16 & F791W &  1.4e-16 &      4.4e-14  \\
UGC7115 &         & F555W &  3.4e-17 & F814W &  3.4e-17 &      6.6e-15  \\
NGC4261 & 3C270   & F547W &  3.8e-18 & F791W &  1.0e-17 &      3.2e-14  \\
NGC4335 &         & F555W & $<$2.5e-17 & F814W & $<$3.0e-17 &  8.6e-15  \\
NGC4374 & 3C272.1 & F547W &  6.2e-17 & F814W &  7.3e-17 &      3.9e-14 \\
NGC4486 & 3C274   & F555W &  6.4e-16 & F814W &  3.2e-16 &      6.0e-14 \\
NGC5127 &         & F555W & $^{\dag}$ & F814W & $^{\dag}$ &    1.6e-15\\
NGC5141 &         & F555W & $<$9.5e-18 & F814W & $<$1.4e-17 &  6.5e-15  \\
NGC5490 &         & F555W & $<$6.7e-19 & F814W & $<$4.0e-18 &  6.1e-15  \\
NGC7052 &         & F547W & $^{\dag}$ & F814W  & $^{\dag}$  &  1.4e-14\\
UGC12064& 3C449   & $^{\ddag}$ & -    & F702W & 3.1e-17 &      5.8e-15  \\
NGC7626 &         & F555W & $<$3.7e-18 & F814W & $<$1.1e-17 &  7.5e-15  \\
\enddata
\tablecomments{The optical and {\HalphaNII} core fluxes from WFPC2 observations. 
Col.(3)-(6): $V$- and $I$-band filter NOS flux in $\ergscmA$. Upper
limits are given for sources in which no optical core was detected
(i.e., FWHM $>0.08''$; see Section~\ref{s:opticalcores}). Notes:
$^{\dag}$: nucleus hidden from view by dust. $^{\ddag}$: no WFPC2
$V$-band observation available. Col.(7): {\HalphaNII} core flux in
$\ergscm$.}
\end{deluxetable}


\begin{deluxetable}{llllll}
\tablewidth{0pt}
\tablecaption{Core flux correlations\label{t:nucflux}}
\tablehead{
\colhead{Flux 1} & \colhead{Flux 2} & \colhead{Cox} & \colhead{Kendall} & 
\colhead{Schmitt(slope,intercept)} & \colhead{LR(slope,intercept)} \\
\colhead{(1)} & \colhead{(2)} & \colhead{(3)} & \colhead{(4)} & \colhead{(5)} & 
\colhead{(6)} \\
}
\startdata
VLBA core       & {\HalphaNII}     & 0.0001 & 0.0001 & 0.58(0.09) 0.45(2.19) & 0.66(0.08) 2.08 \\ 
Host magn.      & {\HalphaNII}     & 0.0212 & 0.0783 & & \\
Radio tot       & {\HalphaNII}     & 0.0060 & 0.1730 & & \\
5 GHz core      & {\HalphaNII}     & 0.0232 & 0.0240 & & \\ 
VLA core        & {\HalphaNII}     & 0.0002 & 0.0004 & 0.64(0.13) 1.40(3.27) & 0.63(0.10) 1.38 \\ 
VLBA core       & NOS $I$          & 0.0002 & 0.0006 & & \\
VLBA core       & NOS $V$          & 0.0003 & 0.0009 & & \\
NOS $I$         & {\HalphaNII}     & -      & 0.0035 & & \\
NOS $V$         & {\HalphaNII}     & -      & 0.0046 & & \\
VLBA ext.       & {\HalphaNII}     & 0.0038 & 0.0044 & 0.61(0.14) -1.27(3.38) & 0.59(0.13) 0.87 \\ 
VLA ext.        & {\HalphaNII}     & 0.0435 & 0.1107 & & \\
VLBA core       & VLBA ext.        &$<0.0001$ &$<0.0001$ & & \\ 
\enddata
\tablecomments
{Statistical significance of various flux-flux correlations and the
linear regression fits to the logarithm of the fluxes. Col.(1)-(2):
the pair of fluxes under consideration. Flux 1 and 2 are the
independent and dependent variable, respectively. Fluxes are in
$\ergscm$ or $\ergscmHz$; host magnitude is the apparent photographic
magnitude (paper I). The abbreviation ext.\ refers to extended
flux. Col.(3)-(4): probability no correlation is present between the
two quantities using Cox's proportional hazard model. Col.(5)-(6): the
slope and intercept of a linear regression fit using Schmitt's method
and with Kaplan-Meier residuals (Isobe, Feigelson \& Nelson 1986). The
errors on the parameters are given between parentheses.}
\end{deluxetable}


\begin{deluxetable}{llllll}
\tablewidth{0pt}
\tablecaption{Core luminosity correlations\label{t:nuclum}}
\tablehead{
\colhead{Lum. 1} & \colhead{Lum. 2} & \colhead{Cox} & \colhead{Kendall} & 
\colhead{Schmitt(slope,intercept)} & \colhead{LR(slope,intercept)} \\
\colhead{(1)} & \colhead{(2)} & \colhead{(3)} & \colhead{(4)} & \colhead{(5)} & 
\colhead{(6)} \\
}
\startdata
VLBA core    & {\HalphaNII}   & $<0.0001$ & 0.0001 &  0.68(0.11)  19.71(3.24) &  0.69(0.11)    19.61 \\
Host magn.   & {\HalphaNII}   &    0.3138 & 0.3465 & -0.27(0.19)  34.28(3.88) & -0.27(0.19)   34.29 \\  
Radio tot.   & {\HalphaNII}   &    0.1049 & 0.4171 &  0.26(0.21)  31.72(6.52) &  0.26(0.19)   31.59 \\
5 GHz core   & {\HalphaNII}   &    0.0523 & 0.0734 &  0.47(0.26)  25.81(7.58) &  0.48(0.23)   25.46 \\
VLA core     & {\HalphaNII}   &    0.0001 & 0.0002 &  0.65(0.12)  20.39(3.49) &  0.65(0.12)   20.58 \\
VLBA core    & NOS $I$        &    0.0001 & 0.0003 &  -                       &  1.18(0.25)   -8.45 \\   
VLBA core    & NOS $V$        &    0.0004 & 0.0007 &  -                       &  1.47(0.35)  -17.27 \\ 
NOS $I$      & {\HalphaNII}   &    -      & 0.0057 &  -                       &  0.55(0.17)   25.31$^{\dagger}$ \\
NOS $V$      & {\HalphaNII}   &    -      & 0.0062 &  -                       &  -                  \\
VLBA ext.    & {\HalphaNII}   &    0.0006 & 0.0013 &  0.60(0.09) 22.38(2.62)  &  0.60(0.11)   22.63 \\
VLA  ext.    & {\HalphaNII}   &    0.0881 & 0.0635 &  -                       &  -                  \\
VLBA core    & VLBA ext.      & $<0.0001$ & 0.0001 & & \\ 
{\HalphaNII} & IRAS  12$\micron$ & 0.9483 & 0.3272 & - & - \\
{\HalphaNII} & IRAS  25$\micron$ & 0.1967 & 0.6872 & - & - \\
{\HalphaNII} & IRAS  60$\micron$ & 0.2768 & 0.0351 & - & - \\
{\HalphaNII} & IRAS 100$\micron$ & 0.6445 & 0.9195 & - & - \\
NOS $I$      & IRAS  12$\micron$ & -      & 0.5400 & - & - \\
NOS $I$      & IRAS  25$\micron$ & -      & 0.4550 & - & - \\
NOS $I$      & IRAS  60$\micron$ & -      & 0.1347 & - & - \\
NOS $I$      & IRAS 100$\micron$ & -      & 0.8119 & - & - \\
inclination  & {\HalphaNII}   &    0.1496 & 0.3918 & - & - \\
inclination  & NOS $I$        &    0.2056 & 0.1691 & - & - \\
inclination  & NOS $V$        &    0.1601 & 0.1105 & - & - \\ 
inclination  & NOS $V-I$      &    0.9224 & 0.7538 & - & - \\ 
inclination  & \tiny{res. radio core - {\HalphaNII}} & 0.5136 & 0.6767 & & \\
inclination  & \tiny{res. radio core - NOS $I$}      & 0.3003 & 0.2990 & & \\
$\mu_I$      & {\HalphaNII}   & 0.5209 & 0.5848 & & \\  
\enddata
\tablecomments{
Statistical significance of various luminosity-luminosity
correlations and linear regression fits to the logarithm of the
luminosities. The columns are similar to those in Table~\ref{t:nucflux}. 
All luminosities are in $\ergs$ or $\ergsHz$; host magnitude is the absolute photographic
magnitude (paper I). The abbreviation ext.\ refers to extended
emission, inclination refers to dust-disk inclination, res. refers to residuals from the correlations and $\mu_I$ is the central stellar surface brightness. Notes: $^{\dagger}$: a least squares linear regression method which
takes into account the errors in the measurements of both quantities
is used in this case, because the errors in both measurements are
similar (Press \etal 1992).}
\end{deluxetable}


\begin{deluxetable}{ccrrccr}
\tablewidth{0pt}
\tablecaption{Central emission in UV-bright LINERs\label{t:liners}}
\tablehead{
\colhead{Target} & \colhead{Type} & \colhead{Filter} & 
\colhead{D$_{75}$} & \colhead{NOS} & \colhead{{\HalphaNII}} & \colhead{Radio} \\
 & & & \colhead{(Mpc)} & \colhead{($\ergsHz$)} &
 \colhead{($\ergs$)} & \colhead{($\ergsHz$)} \\
\colhead{(1)} & \colhead{(2)} & \colhead{(3)} & \colhead{(4)} & \colhead{(5)} & 
\colhead{(6)} & \colhead{(7)} \
}
\startdata
NGC 0404 & S0  & F547M &  2.4 & 25.0 & 37.8 & $<25.4^a$ \\
NGC 3031 & Sab & F547M &  3.6 & 25.4 & 39.1 & $26.8^b$  \\
NGC 4569 & Sab & F547M & 12.3 & 27.0 & 40.0 & $<27.1^a$ \\
         &     & F791W &      & -    &      & \\
NGC 4579 & Sb  & F547M & 18.5 & 26.2 & 40.1 & $27.9^c$  \\
         &     & F791W &      & 26.6 &      & \\
NGC 4594 & Sa  & F547M & 12.0 & 26.0 & 39.9 & $28.1^d$  \\
         &     & F814W &      & 26.5 &      & \\
NGC 5055 & Sb  & F606W &  8.3 & -    & 38.5 & $<26.2^a$ \\
NGC 6500 & Sab & F547M & 40.0 & -    & 40.0 & $<29.5^e$ \\
\enddata
\tablecomments{
Central radio, optical continuum and {\HalphaNII} luminosities for the
LINER galaxy sample studied by Maoz \etal (1998). Col.~(2): Hubble
type. Col.~(3): HST/WFPC2 filter for which the optical emission was
determined. Col.~(4): galaxy distances from Maoz \etal
(1998). Col.~(5): the unresolved nuclear optical luminosity. For the
NGC 4569 F791W image and NGC 6500 F547M image no unresolved source
could reliably be detected. The core was saturated in the image of NGC
5055. Col.~(6): central {\HalphaNII} luminosity, typically measured
within a central aperture of a few arcsec$^2$ (Ho, Filippenko \&
Sargent 1997). The fluxes for NGC 3031, NGC 4569 and NGC 5055 were
obtained under non-photometric conditions and might be
underestimated. Col.~(7): (upper
limits to) the core radio luminosity at 1.4 GHz. References for the
radio luminosities $^a$:upper limits from peak flux measurements
obtained from the FIRST radio sky survey which has resolution of FWHM
$ \sim 5''$ (White \etal 1997); $^b$: Turner \& Ho (1994); $^c$: Sadler
\etal (1995); $^d$:Hummel, van der Hulst, \& Dickey (1984); $^e$:upper 
limits from peak flux measurements obtained from the NVSS radio sky survey
which has resolution of FWHM$\sim 45''$ (Condon \etal~1998). The
detections have resolutions of FWHM$\leq 1''$.}
\end{deluxetable}


\begin{deluxetable}{lrrlrr}
\tablewidth{0pt}
\tablecaption{Radio Core Spectral Indices\label{t:spectralindex}}
\tablehead{
\colhead{target} & \colhead{$\alpha_1$} & \colhead{$\alpha_2$} & \colhead{target} & \colhead{$\alpha_1$} & \colhead{$\alpha_2$} \\
\colhead{(1)} & \colhead{(2)} & \colhead{(3)} & \colhead{(1)} & \colhead{(2)} & \colhead{(3)} \\
}
\startdata
NGC 193  &       &       & NGC 4261 & -1.43 & -0.53 \\
NGC 315  & -0.79 & -0.23 & NGC 4335 &       &       \\
NGC 383  &  1.39 &  1.85 & NGC 4374 & -0.13 & -0.07 \\
NGC 541  &       &       & NGC 4486 & -0.73 &  0.04 \\
NGC 741  &       &       & NGC 5127 & -1.60 & -0.95 \\
3C66B    & -0.42 & -0.25 & NGC 5141 & -1.30 & -0.58 \\
NGC 2329 &       &       & NGC 5490 &       &       \\
NGC 2892 &       &       & NGC 7052 & -0.53 & -0.26 \\
NGC 3801 &       &       & 3C449    &       &       \\
NGC 3862 & -0.46 &  0.55 & NGC 7626 &       &       \\
UGC 7115 &       &       &          &       &       \\
\enddata
\tablecomments
{Radio spectral indices (defined by $S_{\nu} \sim
\nu^{-\alpha}$). Col.(2): radio spectral index of VLBA 1670
MHz core (FWHM $\sim 0.01''$) and 5 GHz core (FWHM $\sim 1.4''$ except
NGC 4261: FWHM $\sim 15''$). The typical error due to the VLBA flux
uncertainty is 0.04. The errors for the 5 GHz fluxes are
unknown. Col.(3): radio spectral index VLA 1490 MHz core (FWHM $1.5''-
3.75''$) and 5 GHz core with a typical error of 0.04 due to the VLA
flux uncertainty. Core variability might have a significant effect on
the derived spectral index because the fluxes were not observed at the
same epoch (see Section~\ref{s:accretion}).}
\end{deluxetable}


\begin{deluxetable}{llrrrrr}
\tablewidth{0pt}
\tablecaption{Central emission in LINER and KDC galaxies\label{t:nucfluxliners}}
\tablehead{
\colhead{Target} & \colhead{Type} & \colhead{D$_{75}$} & 
\colhead{Filter} & \colhead{NOS} & \colhead{{\HalphaNII}} & \colhead{Radio} \\
 & & \colhead{(Mpc)} & & \colhead{\tiny ($\ergscmHz$)} &
 \colhead{\tiny ($\ergscm$)} & \colhead{\tiny ($\ergscmHz$)} \\
\colhead{(1)} & \colhead{(2)} & \colhead{(3)} & \colhead{(4)} & \colhead{(5)} & 
\colhead{(6)} & \colhead{(7)} \
}
\startdata
   NGC 474 &         RS0/a &  32.5 &   F814W  & $<$1.25E-27 &   4.28E-15 &$<$5.00E-27 \\
  NGC 3193 &            E2 &  23.2 &   F702W  & $<$1.79E-27 &   6.61E-15 &$<$5.00E-27 \\
  NGC 3226 &    E2/S0.1(2) &  23.4 &   F547M  & $<$1.21E-27 &   3.87E-14 &   3.60E-26 \\
  NGC 3379 &            E0 &   8.1 &   F814W  & $<$3.77E-28 &   3.61E-14 &   7.00E-27 \\
  NGC 3414 &    S0.(1-2)/a &  24.9 &   F814W  & $<$3.19E-27 &$>$6.78E-14 &   5.00E-26 \\
  NGC 3607 &       S0.3(3) &  19.9 &   F814W  & $\dagger$   &   6.83E-14 &   2.60E-26 \\
  NGC 3608 &            E1 &  23.4 &   F814W  & $<$1.75E-27 &   7.88E-15 &$<$5.00E-27 \\
  NGC 4036 &     S0.3(8)/a &  24.6 &   F547M  & $<$6.31E-28 &   1.24E-13 &   2.90E-26 \\
  NGC 4111 &       S0.1(9) &  17.0 &   F547M  & $<$1.86E-27 &   1.99E-13 &   2.30E-26 \\
  NGC 4143 &     S0.1(5)/a &  17.0 &   F606W  &    3.16E-27 &   4.71E-14 &   6.70E-26 \\
  NGC 4203 &       S0.2(1) &   9.7 &   F814W  & $<$3.89E-27 &   6.89E-14 &   1.25E-25 \\
  NGC 4278 &            E1 &   9.7 &   F814W  &    1.46E-27 &   3.46E-13 &   3.51E-24 \\
  NGC 4494 &            E1 &   9.7 &   F814W  & $<$5.87E-27 &$<$9.72E-15 &$<$5.00E-27 \\
  NGC 4550 &    E7/S0.1(7) &  16.8 &   F814W  & $<$1.70E-27 &$>$1.35E-14 &$<$1.00E-26 \\
  NGC 4589 &            E2 &  30.0 &   F814W  & $<$1.15E-27 &   3.25E-14 &   2.10E-25 \\
  NGC 4762 &          S0.1 &  16.8 &   -      &   -         &   3.59E-15 &$<$5.00E-27 \\
  NGC 5322 &            E4 &  31.6 &   F814W  & $\dagger$   &   2.14E-14 &   2.00E-25 \\
  NGC 5353 &    S0.1(7)/E7 &  37.8 &   -      &   -         &   2.55E-14 &   3.40E-25 \\
  NGC 5354 &  S0 (spindle) &  32.8 &   -      &   -         &   1.11E-14 &   8.10E-26 \\
  NGC 5485 &       S0.3(2) &  32.8 &   -      &   -         &   5.98E-15 &$<$1.00E-26 \\
  NGC 5631 &     S0.3(2)/a &  32.7 &   -      &   -         &   2.58E-14 &$<$5.00E-27 \\
  NGC 5813 &            E1 &  28.5 &   F814W  & $<$4.67E-29 &$>$1.44E-14 &   2.10E-26 \\
  NGC 5982 &            E3 &  38.7 &   F814W  & $<$2.90E-28 &   6.70E-15 &$<$5.00E-27 \\
\hline
  NGC 4278 &            E1 &   9.7 &   F814W  &    1.46E-27     &   3.46E-13 &   3.51E-24 \\
  NGC 4552 &            E  &  13.8 &   F814W  &    2.21E-28$^1$ &   2.95E-14 &   1.08E-24$^a$ \\
  NGC 7192 &            E  &  36.8 &   F814W  & $<$6.08E-29$^2$ &   3.02E-13 &   -        \\
  IC 1459  &            E  &  22.1 &   F814W  &    2.65E-27$^3$ &   7.76E-14 &   1.20E-23$^b$ \\
\enddata
\tablecomments{
Central radio, optical continuum and {\HalphaNII} luminosities for the
LINER galaxy sample divided by a horizontal line from similar data for
the KDC sample. NGC 4278 is part of both samples. The
samples are discussed in Section~\ref{s:fr1sandllagns}. Col.~(2)-(3):
Hubble type and distance from Ho \etal (1997) for the LINER sample and
from Carollo \etal (1997a,b) for the KDC sample. Col.~(4)-(5) Nuclear
optical source filter and flux for galaxies with HST/WFPC2
observations. Note: $\dagger$: central dust inhibits a reliable
estimate of the NOS flux. The NOS measurements for the KDC sample are
taken from Carollo~\etal (1997a,b). Col.~(6): central {\HalphaNII}
luminosity, typically measured within $2''$x$4''$ for the LINER
sample. Lower limits are due to non-photometric observing
conditions. References for the KDC sample: $^1$: Cappellari \etal
(1999); $^3$: Macchetto \etal (1996); $^4$: Verdoes Kleijn \etal
(2000). Col.~(7): the VLA core radio flux at 5 GHz (FWHM $\sim 5''$)
taken from Wrobel \& Heeschen (1991) for the LINER sample. References
to 1.4 GHz observations for the KDC sample: $^a$: Becker \etal (1995);
$^b$: Condon \etal (1998).}
\end{deluxetable}

\begin{deluxetable}{lcllll}
\tablewidth{0pt}
\tablecaption{General properties\label{t:generalproperties}}
\tablehead{
\colhead{UGC} & \colhead{Type} & \colhead{D$_{75}$}
& \colhead{M$_{\rm p}$} & \colhead{Log L$_{\rm radio}$} & \colhead{Comment} \\
 & & \colhead{(Mpc)}
& \colhead{(mag)} & \colhead{(WHz$^{-1}$)} & \\
\colhead{(1)} & \colhead{(2)} & \colhead{(3)} & \colhead{(4)} & \colhead{(5)} & 
\colhead{(6)} \\}
\ifsubmode\renewcommand{\arraystretch}{0.68}\fi
\startdata
07115 & E    & 90.5 & -20.3 & 23.85 & \nl
12064 & E-S0 & 68.3 & -19.9 & 24.29 & 3C 449 \nl
\enddata
\tablecomments{General properties of UGC 7115 and UGC 12064. The general 
properties of the other sample galaxies were presented in paper
I. Col.~(1): UGC number. Col.~(2): Hubble type from Condon \&
Broderick (1988). Col.~(3): distances from Faber \etal (1989) or if
not available directly from recession velocity and $H_0$=75
kms$^{-1}$Mpc$^{-1}$. Col.~(4)-(5): photographic magnitude and total
spectral radio luminosity at 1400 MHz from Condon \& Broderick (1988).
Col.~(6): 3C catalog number.}
\end{deluxetable}


\begin{deluxetable}{llr}
\tablewidth{0pt}
\tablecaption{HST/WFPC2 Observational Log \label{t:obs}}
\tablehead{
\colhead{Target} & \colhead{Filter} & \colhead{Exposure Time} \\
 & & \colhead{(s)} \\
\colhead{(1)} & \colhead{(2)} & \colhead{(3)} \\
}
\startdata
UGC 7115 & F555W  & 1000 \\ 
UGC 7115 & F814W  &  800 \\ 
3C 449 & F702W  &  560 \\
3C 449 & LRF680 &  600 \\
\enddata
\tablecomments{Observational log of UGC 7115 and 3C 449. 
Col.~(2):
  HST/WFPC2 filter name Col.~(3): total exposure time of each
  observation.}
\end{deluxetable}



\end{document}